%% file: main.tex
\documentclass[sigconf, nonacm]{acmart}

\usepackage{xspace}
\usepackage{ulem}

\usepackage{bm}
\usepackage{setspace}
\usepackage{enumitem}
\usepackage{comment}
\usepackage{amsmath}
\usepackage{amsthm}
\usepackage[lined,linesnumbered,boxed,vlined,ruled]{algorithm2e}
\usepackage{float}
\usepackage{colortbl}
\usepackage{subcaption}
\usepackage{caption}

\newcommand\vldbdoi{XX.XX/XXX.XX}
\newcommand\vldbpages{XXX-XXX}
\newcommand\vldbvolume{14}
\newcommand\vldbissue{1}
\newcommand\vldbyear{2020}
\newcommand\vldbauthors{\authors}
\newcommand\vldbtitle{\shorttitle} 
\newcommand\vldbavailabilityurl{URL_TO_YOUR_ARTIFACTS}
\newcommand\vldbpagestyle{plain} 

\newcommand\revise[1]{{\color{black}#1}}
\newcommand\yuxing[1]{{\color{black}#1}}

\graphicspath{{./images/}}

\newcommand{\AQEs}{\textsf{AQEs}\xspace}
\newcommand{\AQETune}{\textsf{AQETuner}\xspace}
\newcommand{\HSPE}{\textsf{HSPE}\xspace}

\theoremstyle{definition}

\definecolor{mygrey}{RGB}{230,230,240}
\begin{document}
\title{\AQETune: Reliable Query-level Configuration Tuning for Analytical Query Engines}


\author{Lixiang Chen}
\affiliation{%
  \institution{East China Normal University}
  \institution{\& ByteDance}
}
\email{lxchen@stu.ecnu.edu.cn}

\author{Yuxing Han}
\authornote{Corresponding authors.}
\affiliation{%
  \institution{ByteDance}
}
\email{hanyuxing@bytedance.com}

\author{Yu Chen}
\affiliation{%
  \institution{East China Normal University}
  \institution{\& ByteDance}
}
\email{yuchen198@stu.ecnu.edu.cn}

\author{Xing Chen}
\affiliation{%
  \institution{ByteDance}
}
\email{chenxing.xc@bytedance.com}

\author{Chengcheng Yang}
\authornotemark[1]
\affiliation{%
  \institution{East China Normal University}
}
\email{ccyang@dase.ecnu.edu.cn}

\author{Weining Qian}
\affiliation{%
  \institution{East China Normal University}
}
\email{wnqian@dase.ecnu.edu.cn}

\begin{abstract}
Modern analytical query engines (\AQEs) are essential for large-scale data analysis and processing.
These systems usually provide numerous query-level tunable knobs that significantly affect individual query performance.
While several studies have explored automatic DBMS configuration tuning, they have several limitations to handle query-level tuning. 
Firstly, they fail to capture how knobs influence query plans, which directly affect query performance.
Secondly, they overlook query failures during the tuning processing, resulting in low tuning efficiency. 
Thirdly, they struggle with cold-start problems for new queries, leading to prolonged tuning time.
To address these challenges, we propose \AQETune, a novel Bayesian Optimization-based system tailored for \textit{reliable} query-level knob tuning in \AQEs. 
\AQETune first applies the attention mechanisms to jointly encode the knobs and plan query, effectively identifying the impact of knobs on plan nodes.
Then, \AQETune employs a dual-task Neural Process to predict both query performance and failures, leveraging their interactions to guide the tuning process.
Furthermore, \AQETune utilizes Particle Swarm Optimization to efficiently generate high-quality samples in parallel during the initial tuning stage for the new queries.
Experimental results show that \AQETune significantly outperforms existing methods, reducing query latency by up to \yuxing{23.7\%} and query failures by up to \yuxing{51.2\%}.
\end{abstract}

\maketitle

\vspace{-2mm}
\pagestyle{\vldbpagestyle}
\begingroup\small\noindent\raggedright\textbf{PVLDB Reference Format:}\\
\vldbauthors. \vldbtitle. PVLDB, \vldbvolume(\vldbissue): \vldbpages, \vldbyear.\\
\href{https://doi.org/\vldbdoi}{doi:\vldbdoi}
\endgroup
\begingroup
\renewcommand\thefootnote{}\footnote{\noindent
This work is licensed under the Creative Commons BY-NC-ND 4.0 International License. Visit \url{https://creativecommons.org/licenses/by-nc-nd/4.0/} to view a copy of this license. For any use beyond those covered by this license, obtain permission by emailing \href{mailto:info@vldb.org}{info@vldb.org}. Copyright is held by the owner/author(s). Publication rights licensed to the VLDB Endowment. \\
\raggedright Proceedings of the VLDB Endowment, Vol. \vldbvolume, No. \vldbissue\ %
ISSN 2150-8097. \\
\href{https://doi.org/\vldbdoi}{doi:\vldbdoi} \\
}\addtocounter{footnote}{-1}\endgroup


\ifdefempty{\vldbavailabilityurl}{}{
\vspace{.3cm}
\begingroup\small\noindent\raggedright\textbf{PVLDB Artifact Availability:}\\
The source code, data, and/or other artifacts have been made available at \url{https://gitfront.io/r/ml4tuning/aoR8ReCrj4a7/aqetuner/}.
\endgroup
}


\input{sections/introduction}

\input{sections/background}

\input{sections/overview}

\input{sections/featurization}

\input{sections/np_tune}
\input{sections/pso_tune}
\input{sections/evaluation}

\input{sections/related_work}

\input{sections/conclusion}

\balance

\newpage


\normalem
\bibliographystyle{ACM-Reference-Format}
\bibliography{ref}

\end{document}

%% file: sections/introduction.tex
\section{Introduction}







\begin{figure}[htb]
\captionsetup[subfigure]{skip=-0.1em} 
    \centering \label{fig: motivation}
    \subfloat[\footnotesize Knobs Distribution]{
        \includegraphics[height=2.7cm]{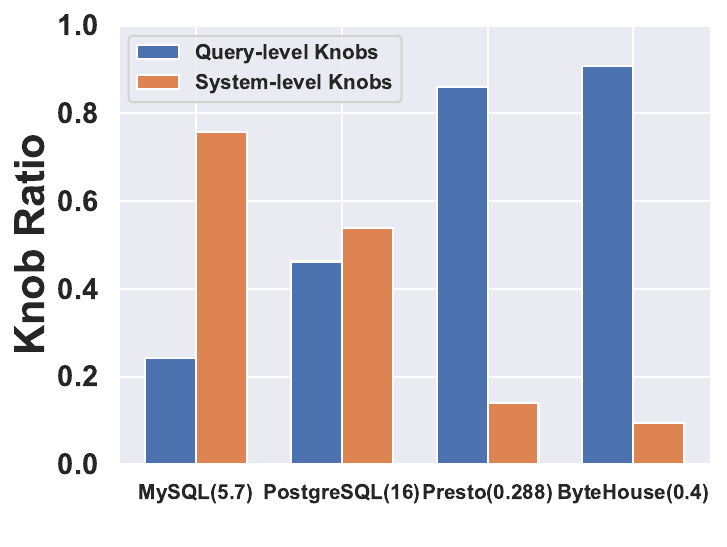}
       
        \label{fig:ob1}
    }
    \hfill
    \subfloat[\footnotesize Increasing Query-Level Knobs]{
        \includegraphics[height=2.7cm]{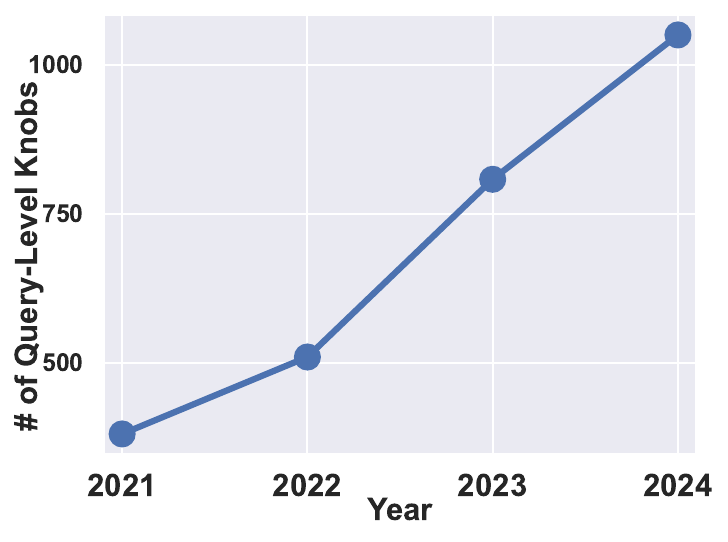}
        \label{fig:ob2}
    }   \\ \ \ \ \ 
    \subfloat[\footnotesize Heatmap of Query Failures]{
        \includegraphics[width=0.2\textwidth]{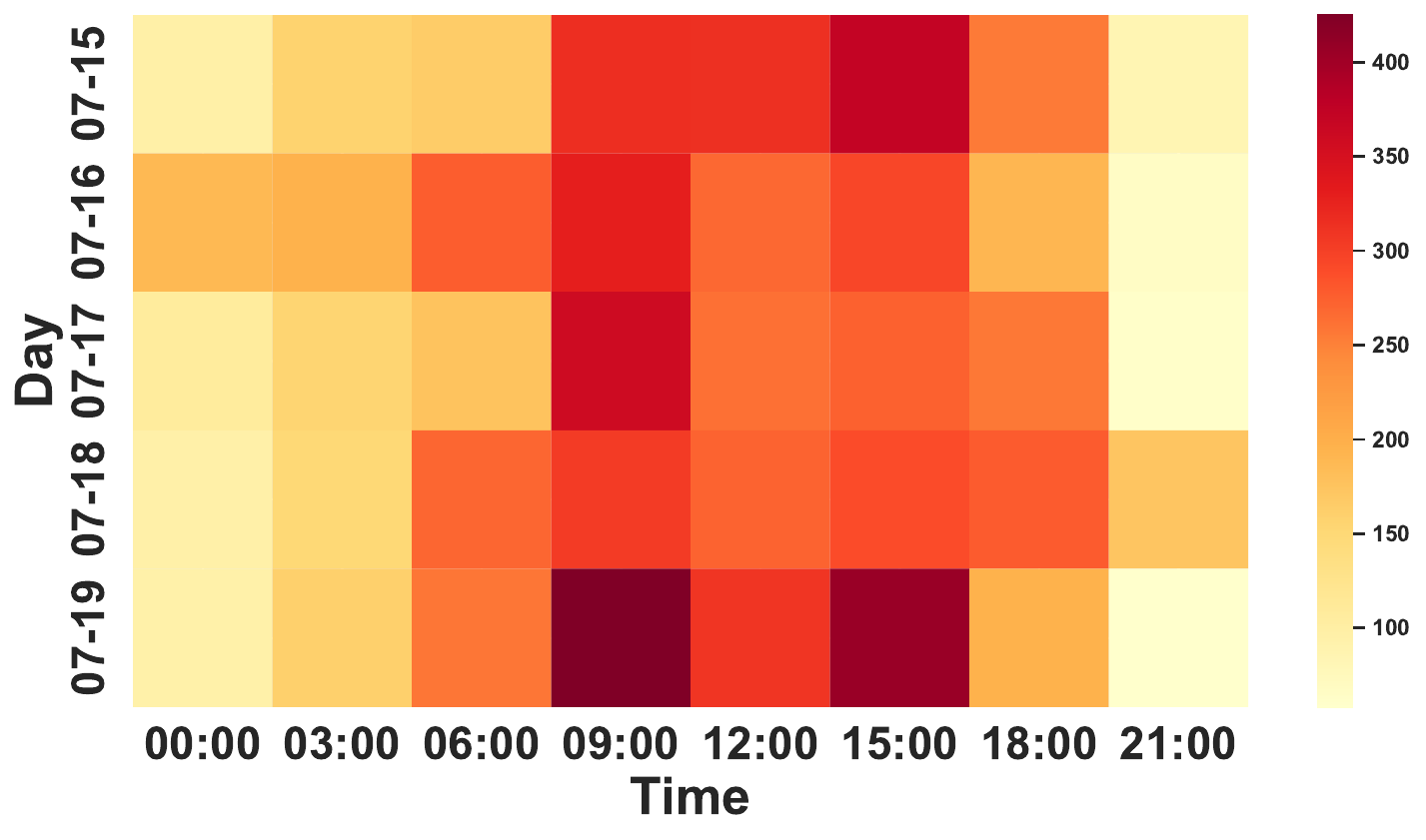}
        \label{fig:ob3}
    }
    \hfill
    \subfloat[\footnotesize Categories of Error Messages]{
        \includegraphics[width=0.18\textwidth]{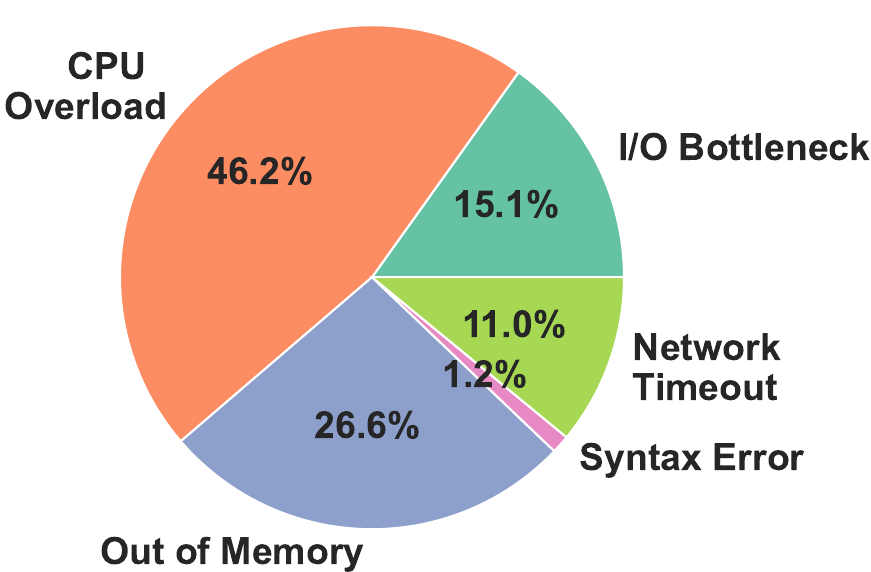}
        \label{fig:ob4}
    }
    \vspace{-1em}
    \caption{Motivating Examples}
    \vspace{-3em}
\end{figure}

Modern analytical query engines (\AQEs), including Snowflake~\cite{dageville2016snowflake}, Presto~\cite{sun2023presto}, and ByteHouse~\cite{han2024bytecard}, are integral to business intelligence and decision-making across industries.
For example, ByteHouse handles 200 million queries every day, processing multiple exabytes of data to meet the analytical demands of platforms such as TikTok.
The performance of \AQEs relies heavily on numerous tunable configurations (or knobs). 
However, knob tuning is an NP-hard problem~\cite{sullivan2004nphard} and requires a deep understanding of query optimization. Even experienced DBAs often struggle to fully grasp the impact of configurations on every individual query execution.

To reduce the manual tuning efforts of DBAs, recent approaches have sought to automate the knob tuning via Machine Learning (ML) techniques, including Bayesian Optimization (BO)~\cite{van2017ottertune,zhang2021restune,cereda2021cgptuner,duan2009itune,zhang2022onlinetune} and Deep Reinforcement Learning (DRL)~\cite{li2019qtune,zhang2019cdbtune,ge2021watuning,cai2022hunter}. 
BO offers a theoretically-sound method for exploring the configuration space with improvement guarantees, while DRL utilizes Deep Deterministic Policy Gradient (DDPG)~\cite{ddpg},  a model-free and actor-critic algorithm that operates on continuous action (i.e., configuration) spaces.
Most of these approaches target specific workloads and optimize the overall throughput and latency of transactional databases.
%


However, these approaches have limitations for tuning \AQEs.
On the one hand, \AQEs usually have different knob distributions compared to transactional databases.
The database knobs can be broadly classified into two main categories: system-level knobs, which control the overall system behavior (e.g., buffer pool size and log file size~\cite{van2017ottertune}), and query-level knobs, which optimize individual query execution (e.g., runtime operator parallelism~\cite{mehta1995managing} and execution strategy option~\cite{laptev2022smarter}).
It is often observed that \AQEs expose a higher proportion of query-level knobs.
Fig.~\ref{fig:ob1} shows the ratio of knob categories in different transactional databases and \AQEs, with ByteHouse and Presto having more query-level knobs than PostgreSQL and MySQL.
Moreover, Fig.~\ref{fig:ob2} highlights the rapid increase in query-level knobs in ByteHouse as the system evolves.
This reflects the need for \AQEs to offer more fine-grained tuning  knobs~\cite{bytehouseConf, prestoConf, snowflakeConf} to handle complex analytical queries that involve aggregations, joins, and window functions on large datasets~\cite{sun2023presto,dageville2016snowflake}.

On the other hand, analytical engines frequently encounter query failures due to the complexity of query processing and the substantial data volumes involved. 
Fig.~\ref{fig:ob3} shows the heatmap of query failures throughout one workday in a ByteHouse cluster, while Fig.~\ref{fig:ob4} categorizes the corresponding error messages.
Most failures are attributed to suboptimal query-level knob configurations, such as inadequate CPU/memory/storage resource allocations, improper parallelism settings, and poor execution strategies. 

Given these tuning requirements, tuning \AQEs should meet two criteria, which existing methods struggle to achieve:
1) \uline{Query-Level Tuning}: queries must be optimized using query-level knobs.
2) \uline{Reliability}: methods should ensure that query failure rates do not increase during the tuning process.
Specifically, the key challenges of tuning the analytical engines can be summarized as follows:

\textbf{Jointly Encoding the Knobs and Query Plans (C1).}
Joint representation learning for knobs and queries should capture the impact of knobs on the query execution to enhance downstream tuning tasks.
However, encoding them in a shared space poses a complex challenge.
Since query plans are closely related to the query execution, we propose to encode the query plan rather than the plain query text. 
To this end, two critical issues must be addressed: First, how to effectively preserve the hierarchical structure of the tree-structured query plans.
Second, how to model the correlation between the query-level knobs and plan nodes, considering knobs impact different nodes in different ways. 
\textbf{Dual-task Knob Tuning (C2).} 
When performing query-level tuning, both query performance and reliability should be considered, as improper knob settings would increase query failures and reduce tuning efficiency.
Modeling these two targets is challenging because query performance and reliability are inherently interdependent.
For instance, a knob that improves query performance might increase resource consumption, potentially leading to query failures.
As a result, modeling both tasks separately is ineffective, as it overlooks important correlations.
Moreover, integrating encodings from representation learning further complicates the task. 

\textbf{Cold-start Problem (C3). } 
ML-based tuning systems often face the cold-start issue when encountering new queries without prior knowledge~\cite{cai2022hunter}.
Collecting high-quality observation data in the early tuning stage is crucial to accelerate the subsequent tuning process. 
In the absence of such data, the system might experience suboptimal exploration of promising regions, leading to prolonged tuning times.
Moreover, the ability to quickly collect such data is essential to maintain overall tuning efficiency.



\textbf{Our Approach.} To address the above challenges, we propose \AQETune, a novel system for reliable query-level tuning in \AQEs.
\AQETune comprises three main components: knob-plan encoder (addressing C1),  dual-task predictor (addressing C2), and warm starter (addressing C3).
Specifically, the knob-plan encoder first utilizes the self-attention mechanism~\cite{ashish2017attention} to independently encode the query plan and knobs, capturing each of their internal correlations. 
For tree-structured query plans, we propose a hierarchical spectral position encoding to preserve the global structure information 
between plan nodes. 
Then, 
the encoder utilizes the cross-attention mechanism~\cite{gheini2021crossattn} to jointly synergize the encodings of query plans and knobs into a shared space, which further captures each knob's influence on the plan nodes.
To improve computational efficiency, the cross-attention scores are only computed between knobs and plan nodes that are correlated.

The dual-task predictor leverages Neural Process (NP)~\cite{garnelo2018nps,garnelo2018condnp,kim2019attentivenp} to predict both query performance and reliability with the joint knob-plan encodings.
The main characteristic of NP is that it can combine the strengths of neural networks with stochastic processes and allow the efficient fitting of observed data. 
By using NPs, we can naturally leverage the joint encodings within the neural network's embedding space.
However, the vanilla NP lacks the capacity to capture the task correlations and struggles to express the complexity of both tasks.
To enhance model expressiveness, we introduce two intra-task latent variables to handle information related to the prediction tasks of performance and reliability, respectively.  
Moreover, cross-attention is employed to enable the model to selectively focus on the observation data that are most relevant to each task.
To further exploit the inherent correlations between tasks, we introduce a cross-task latent variable to capture the shared information across both tasks.
Subsequently, this variable is used to complement the two intra-task variables with a gating mechanism~\cite{gu2020improving}, ensuring that it effectively contributes to the final predictions. 

The warm starter employs the Particle Swarm Optimization (PSO)~\cite{shami2022particle} to efficiently generate high-quality initial observation data.
PSO uses multiple search agents, each selecting samples within the configuration space iteratively. 
These agents are guided by both the global optimum across all agents and the local optimum of each individual agent.
The search process is further accelerated by leveraging the \AQEs' parallel execution capability, allowing each agent to search in parallel.

In summary, we make the following contributions:
\begin{itemize}[topsep=0pt]
    \item We propose \AQETune, a novel ML-based tuning system for reliable query-level tuning. To our knowledge, this is the first approach that focuses on query-level tuning for analytical query engines.
    \item We solve the challenge of jointly encoding knobs and query plans with various attention techniques.
    \item We address prediction for performance and reliability with a dual-task Neural Process, leveraging the correlation between the tasks to improve accuracy.
    \item We resolve the cold-start issue by applying the Particle Swarm Optimization to generate high-quality samples in the initial tuning phase, ensuring a diverse and efficient exploration of the search space.
    \item   We conduct extensive experiments on both synthetic and real-world workloads. Experimental results show that \AQETune improves the query latency by up to \yuxing{23.7\%} and reduces the query failure rate by up to \yuxing{51.2\%} compared to state-of-the-art methods.
    \end{itemize}



%% file: sections/background.tex
\vspace{-2em}
\section{Preliminary}
In this section, we introduce basic concepts in \AQEs, including the query plan and query-level knob, and provide a brief overview of BO-based automatic tuning for database systems. We then clarify related terminology and formulate the query-level tuning problem for analytical engines.




\begin{figure}[htb]
    \centering \captionsetup[subfigure]{skip=-0.1em}
    \vspace{-1.4em}
    \subfloat[The Most Complex TPC-C Query]{
    \makebox[0.24\textwidth]{\includegraphics[height=0.22\textwidth]{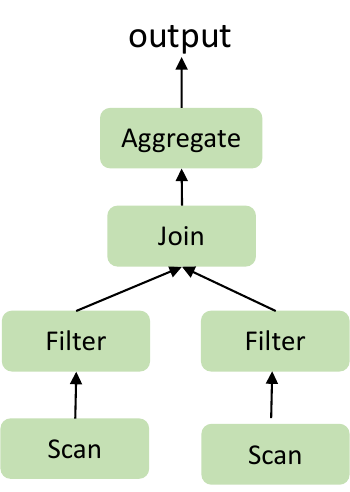}
        \label{fig:tpcc}
    }
    }
    \hfill
    \subfloat[A TPC-DS Query (Q99)]{
        \includegraphics[height=0.22\textwidth]{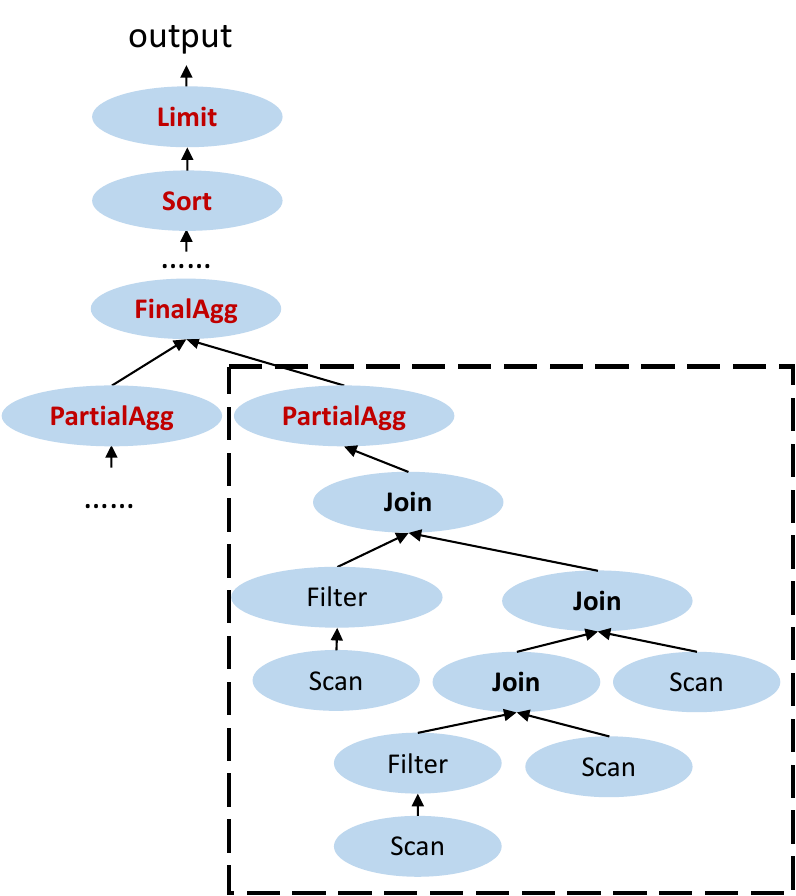}
        \label{fig:tpcds}
    }
    \vspace{-1em}
    \caption{Comparison of Query Plans}
    \vspace{-1.5em}
\label{fig: plan_comp}    
\end{figure}

\subsection{Query Plan}

A query plan in a database is a structured representation of how a query will be executed, usually depicted as a tree-structured Directed Acyclic Graph (DAG).
In this DAG, each node represents an \textit{operator} that performs tasks like table scans, record filtering, joins, or aggregations, while the edges indicate the data flow between these operators.
To effectively model an analytical query, it is preferable to model the query plan directly rather than the query text since the plan captures the runtime behavior of the engine.

Analytical queries are generally more complex than transactional queries, along with their query plans. This is because they often involve numerous complex operations such as joins across several tables, subqueries, data groupings, and advanced functions such as window functions.
In contrast, transactional queries usually consist of simpler CRUD (Create, Read, Update, Delete) operations, which primarily focus on row-level locking and transactional integrity.

Figure~\ref{fig: plan_comp} compares the query plans for the most complex transactional query from the TPC-C benchmark~\cite{tpcc} and a typical analytical query from TPC-DS benchmark~\cite{tpcds}. 
The first plan is generated by PostgreSQL, while the second is produced by ByteHouse.
The key differences are:
1)  The \textit{longest path} in the DAG-structured plan of the analytical query is significantly longer than that of the transactional query.
2) The number and variety of nodes (i.e., operators) in the analytical query plan are much greater.
For example, the plan of TPC-DS query Q14 in ByteHouse has nearly $150$ plan nodes, significantly more complex than TPC-C queries.



\subsection{Query-Level Knobs}
\AQEs offer various query-level tuning knobs that enable users to control and optimize the runtime operator of each individual query.
These knobs are vital for enhancing query performance, especially in distributed environments where efficient data movement across multiple computing instances is critical.
They enable fine-grained tuning, such as resource allocation and runtime settings for specific queries, without changing the overall system configuration.
Common query-level knobs include memory allocation, which sets the memory available to execute an operator; the degree of parallelism, which determines the number of parallel processes or threads used to execute specific operators; and execution strategy options, such as deciding whether to shuffle data before or after the aggregation operator to minimize data transfers~\cite{liu2014automating}.

These knobs can be either categorical/discrete (e.g., turning an execution option on or off) or numerical/continuous (e.g., setting the maximum memory for a specific operator).
Unlike discrete knobs, continuous knobs can take any value within a defined range, which significantly expands the search space and makes finding the optimal settings more challenging.
This complexity is further increased by the interdependencies between query-level knobs, where the optimal value of one knob might shift with small changes on another one.
For example, 
the increased parallelism of operators would also lead to more memory contentions among concurrent processes and impact the optimal setting of memory allocation.

\subsection{Related BO-based Knob Tuning}
%
 
The existing Bayesian Optimization (BO) based knob tuning for DBMS~\cite{van2017ottertune,zhang2021restune,cereda2021cgptuner,duan2009itune,zhang2022onlinetune} starts with an initial random set of configurations.
Then, the DBMS is executed under these initial configurations, and performance metrics (e.g., throughput, latency) are observed.
Based on these observation data, a Gaussian Process is built as the surrogate model to estimate the relationship between the configuration and the performance metrics.
An acquisition function selects the next configuration to evaluate, balancing the exploration of new configurations with the exploitation of known promising ones.
This iterative process involves repeatedly evaluating the DBMS with new configuration points, updating the surrogate model, and refining the search for the optimal knobs. 
It continues until a stopping criterion is met, such as reaching a maximum number of iterations.
The configuration that yields the best-observed performance is then chosen for deployment in the production environment.
 
These methods treat knob tuning as a ``black-box'' optimization problem~\cite{kunjir2020black} and rely on limited features (e.g., keyword statistics) extracted from the query text~\cite{zhang2021restune,zhang2022onlinetune}.
However, they ignore the query plans that specify how queries are executed within the query engine.
We argue that incorporating
the query plan could provide valuable query execution information, which further significantly helps to enhance the tuning performance.

\subsection{Problem Definition}

\noindent\textbf{Query Representation Space.}
Given an analytical query $q$, the representation learning derives a vector representation $\omega_q \in \mathbb{R}^d$ based on extracted relevant features. The learned representation can be utilized for the downstream tuning problem.
We denote the set of all possible vector representations as 
$\Omega = \{ \omega_q \}$.

\noindent\textbf{Knob Search Space.} For each query-level knob $k$, whether discrete or continuous, we normalize its domain ${\Theta}$ to a continuous space $[0, 1]$ using min-max uniform scaling.
Given a set of query-level knobs $K=\{k_1, k_2,...,k_n\}$, where each knob $k_i$ has a domain $\Theta_i$, the overall knob search space is represented as $\Theta=\Theta_{1} \times \cdots \times \Theta_{n}$. 
A specific point in this space represents a possible combination of knob values.

\noindent\textbf{Reliability.}
The reliability of an analytical query $q$ under a given configuration $\theta_q \in \Theta$ is determined by its execution status (success or failure) within an execution environment $E$. 
If the query executes successfully (status $0$), the configuration is considered reliable; otherwise, with an execution status of $1$, the configuration is considered unreliable.
The reliability function can be defined as $g:\Omega\: \times\:\Theta \rightarrow \{0, 1\}$. 


\noindent\textbf{Query-Level Performance.}
We define the query-level performance function as $f:\Omega \:\times\:\Theta \rightarrow \mathbb{R}_{>=0}$.
In this paper, we focus on the end-to-end query-level performance metric: the execution time.

\noindent \textbf{Reliable Query-level Tuning.}
The objective of reliable query-level tuning for a given query $q$ is to find the optimal configuration $\theta^*_{q} \in \Theta$ within the knob search space for an analytical query with correct SQL syntax, aiming to minimize the performance metric while avoiding execution failures. 
We formalize this as an optimization problem:
\begin{equation}
\begin{array}{c}
 \underset{\theta}{\arg \min\ } f(\omega_q,  \theta_q), \\
\text{s.t.\ } g(\omega_q,  \theta_q) = 0
\end{array}
\label{equation:constraints}
\end{equation}

%% file: sections/overview.tex
\section{System Overview}
\begin{figure}[htb]
\centering
\vspace{-1.5em}
  \includegraphics[width=1.0\linewidth]{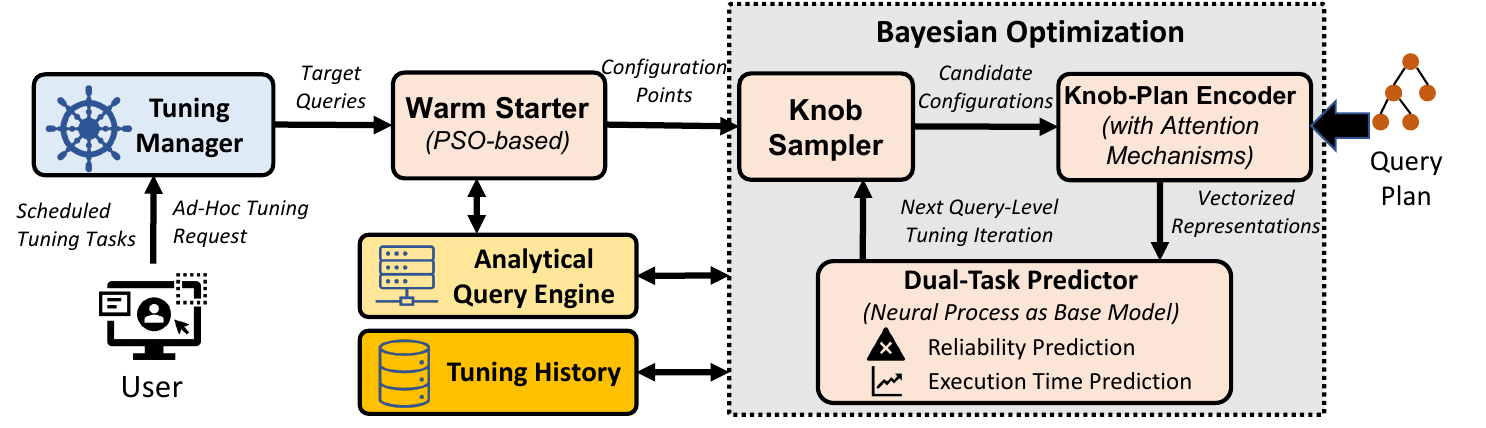}
  \vspace{-2em}
  \caption{\yuxing{Overall Architecture of \AQETune}}
  \vspace{-1.5em}
  \label{fig:framework}
\end{figure}


\noindent \textbf{\yuxing{Components.}}
Figure~\ref{fig:framework} shows the overall architecture of \AQETune, which tunes query-level knobs for the analytical engine using a Bayesian Optimization (BO) framework.
\yuxing{
\AQETune has several critical components to facilitate its knob tuning:
\textit{tuning manager}, which receives target queries from the user and manages the tuning process;
\textit{knob sampler}, which generates candidate configurations at the start of the BO loop;
\textit{knob-plan encoder}, which jointly encodes these configurations and query plans into vectorized representations generated by neural networks with attention mechanisms
(discussed in Section~\ref{section:encode});
\textit{dual-task predictor}, which takes the 
representations as input to predict query reliability $\tilde{g}$ and execution time $\tilde{f}$ and support query-level knob tuning (discussed in Section~\ref{section:np_tuner});
\textit{warm starter}, which collects high-quality observation data using Particle Swarm Optimization (PSO) method and establishes a coarse correlation between the knobs and plan nodes (discussed in Section~\ref{section:warm_starter}).}


Based on dual-task predictions, \AQETune employs the Expected Improvement with Constraints (EIC) ~\cite{gardner2014constraints} as the acquisition function. 
To integrate the reliability constraint, we incorporate the probability of successful execution into expected improvements over the best-observed performance. Specifically, given $f^*$ as the best performance observed so far, the acquisition function is defined as:

\begin{equation}
\begin{array}{c}
    \alpha_{EIC}(f, g) = \mathbb{E}[max(0, \tilde{f} - f^{*})] \times p(\tilde{g}=0).
\end{array}
\end{equation}


\begin{algorithm}[t]
  \small
  \caption{\AQETune Workflow}
  \label{algo: workflow}
   \setlength{\abovecaptionskip}{0.cm}
  \SetAlgoLined
  \KwIn{
  Target queries ${\mathcal{Q}}$; 
  Knob search space ${\bm\Gamma}$;
  Analytical query engine instance $E$;
   Maximum tuning duration $\mathcal{T}$
}
  \KwOut{Best configuration explored for each query}

  Initialize tuning history $\mathcal{H} \gets \emptyset$ \;	
  \While{$\mathcal{T}$ is not reached}
   	{
	$\mathcal{H} \gets$ generateWarmStartSamples($\mathcal{Q}$) \; 
        Consturct a surrogate model $\mathcal{M}$ with $\mathcal{H}$\;
        $\mathcal{C} \gets$ identifyKnobAndPlanCorrelation($\mathcal{H}$, $E$)\;
  \For{each query $q \in {\mathcal{Q}}$} {
  $\mathcal{KS}$ $\gets$ generateCandidateKnobSamples(${\bm\Gamma}$)\;
  Retrieve the query plan $p^{E}$ of $q$ from ${E}$\;

 $\mathcal{X} \gets $ encodeKnobsAndPlan($\mathcal{KS}$, $p^{E}$, $\mathcal{C}$)\;
    
  $\mathcal{B} \gets $ predictReliabilityAndPerformance($\mathcal{X}$, $\mathcal{M}$)\;

${x}^\mathcal{B} \gets$ argmax $\alpha_{EIC}(\mathcal{B})$ \;
$y^\mathcal{B}  \gets$ evaluate corresponding knobs $k^\mathcal{B}$ of ${x}^\mathcal{B}$ on $E$ to obtain true performance of $q$\;
$\mathcal{H} \gets \mathcal{H} \cup (k^\mathcal{B}, y^\mathcal{B} ) $ \;  
    }
    update $\mathcal{M}$ with $\mathcal{H}$\;
    }
  return $k^\mathcal{B}$ with best $y^\mathcal{B}$ for each query in $\mathcal{H}$ \;
\end{algorithm}

\noindent \textbf{Workflow.}
Algorithm~\ref{algo: workflow} outlines the main procedures of \AQETune.
Given the target queries and knob search space, the tuning process is constrained by a duration $\mathcal{T}$.
It starts by initializing the tuning history with the warm starter and constructing an initial surrogate model $\mathcal{M}$ (Lines~3$\sim$4).
Initial configurations are evaluated on the engine $E$ to derive a coarse correlation $\mathcal{C}$ between knobs and plan nodes (Line~5).
For each query, \AQETune samples candidate knobs $\mathcal{KS}$ and retrieves its query plan 
(Lines~6$\sim$8). 
The knob-plan encoder then jointly encodes the query plan and knobs into vectorized encodings $\mathcal{X}$ based on the correlation $\mathcal{C}$ (Line 9). 
The dual-task predictor forecasts both reliability and performance  and selects the most promising encodings ${x}^\mathcal{B}$ 
with the acquisition function
(Lines 10$\sim$11). 
The selected knobs $k^\mathcal{B}$ of $x^\mathcal{B}$ are then evaluated on the analytical engine to collect actual performance metrics (Lines~11$\sim$12), and the results are added to the tuning history (Line~13).
After the tuning iterations of queries are finished, the surrogate model is updated accordingly (Line~15). 
The process iterates until $\mathcal{T}$ is reached, after which the best knob configurations are returned (Line 17).

\yuxing{\noindent \textbf{System Integration.}
To enable seamless integration with existing analytical query engines, \AQETune is designed as an external, non-intrusive service that runs parallel with the running engines.
This design avoids disruption to regular query processing and supports broad compatibility with engines like Presto and ByteHouse.
The integration process follows three key phases:
(1) \textit{Query Plan Extraction}, (2) \textit{Knob Adjustment}, and (3) \textit{Performance Feedback Collection}.
In the \textit{Query Plan Extraction} phase, \AQETune interacts with the engine’s SQL interface to retrieve the query plan using \textsf{EXPLAIN} statements without executing the query.
The plan is then processed by \AQETune’s knob-plan encoder, which jointly encodes the plan with the tunable knobs.
In the \textit{Knob Adjustment} phase, \AQETune generates configuration recommendations, which are applied via external interfaces, such as Presto's \textsf{SET SESSION} command or ByteHouse's query-level \textsf{SETTINGS} syntax.
Finally, during the \textit{Performance Feedback Collection} phase, \AQETune gathers execution feedback (e.g., execution time, execution success/failure) through the client APIs
or system logs. This feedback is fed into the BO loop for iterative refinement of the tuning process. 
}

%% file: sections/featurization.tex
\vspace{-1mm}
\section{Knob-plan Encoder}
\vspace{-1mm}
\label{section:encode}

In this section, we first briefly introduce the attention mechanism and then elaborate on the architecture of the knob-plan encoder. 


\subsection{Basics: Attention Mechanisms}



In recent years, the attention mechanism has become the cornerstone of modern neural networks for sequence modeling~\cite{galassi2020attention, niu2021review}.
This mechanism allows the model to assign different levels of importance to different parts of the input sequence, enabling it to focus dynamically on the most relevant parts when dealing with prediction tasks.
For each token embedding $x\in \mathbb{R}^{d}$ of the input sequence $X\in \mathbb{R}^{n*d}$, where $n$ is the sequence length and $d$ is the embedding dimension,
three vectors Query ($Q$), Key ($K$), and Value ($V$) are constructed by projecting $x$ into three distinct spaces with the learned weight matrices $W_Q \in \mathbb{R}^{d*d_k} $, $W_K \in \mathbb{R}^{d*d_k} $, and $W_V \in \mathbb{R}^{d*d_v}$:
\begin{equation}
Q= x \cdot W_Q, K = x \cdot W_K, V= x \cdot W_V,
\end{equation}
where $d_k$ denotes the dimension of $Q$ and $K$, and $d_v$ denotes the dimension of $V$. 
The dot-product attention score is defined as:
\begin{equation}
\operatorname{Attention}(Q, K, V)=\operatorname{softmax}\left(\frac{Q K^T}{\sqrt{d_k}}\right) V.
\label{eq: attention}
\end{equation}
The output is an encoded representation of the input sequence and
the unnormalized 
score between any two tokens $i$ and $j$ before applying softmax can be expressed in a matrix form:
${A}_{i,j} = \frac{Q_i * K^T_j} {\sqrt{d_k}}$.



There are two variants of attention mechanisms, which are \textit{self-attention}~\cite{ashish2017attention} and \textit{cross-attention}~\cite{gheini2021crossattn}.
Self-attention enables the model to capture relationships between tokens in a sequence, allowing each token to derive a context-aware representation by considering the entire sequence.
Cross-attention, on the other hand, allows information from one sequence to influence the representation of tokens in another. It could generate an output for a sequence that is informed by a different input sequence.

In addition, \textit{positional encodings} are usually added to the input embeddings, allowing the model to recognize the positions of tokens within a sequence. 
Specifically, the Transformer~\cite{ashish2017attention} combines the positional encodings with the attention mechanism.
This, along with residual connections~\cite{wang2017ffn} and a position-wise Feed-Forward Network (FFN), constitutes the \textit{attention block}. 
This structure allows the Transformer to learn and handle complex dependencies in large-scale sequences effectively.

\subsection{Plan Representation} 
\begin{figure}[tb]
\vspace{-3mm}
\centering
  \includegraphics[width=1.0\linewidth]{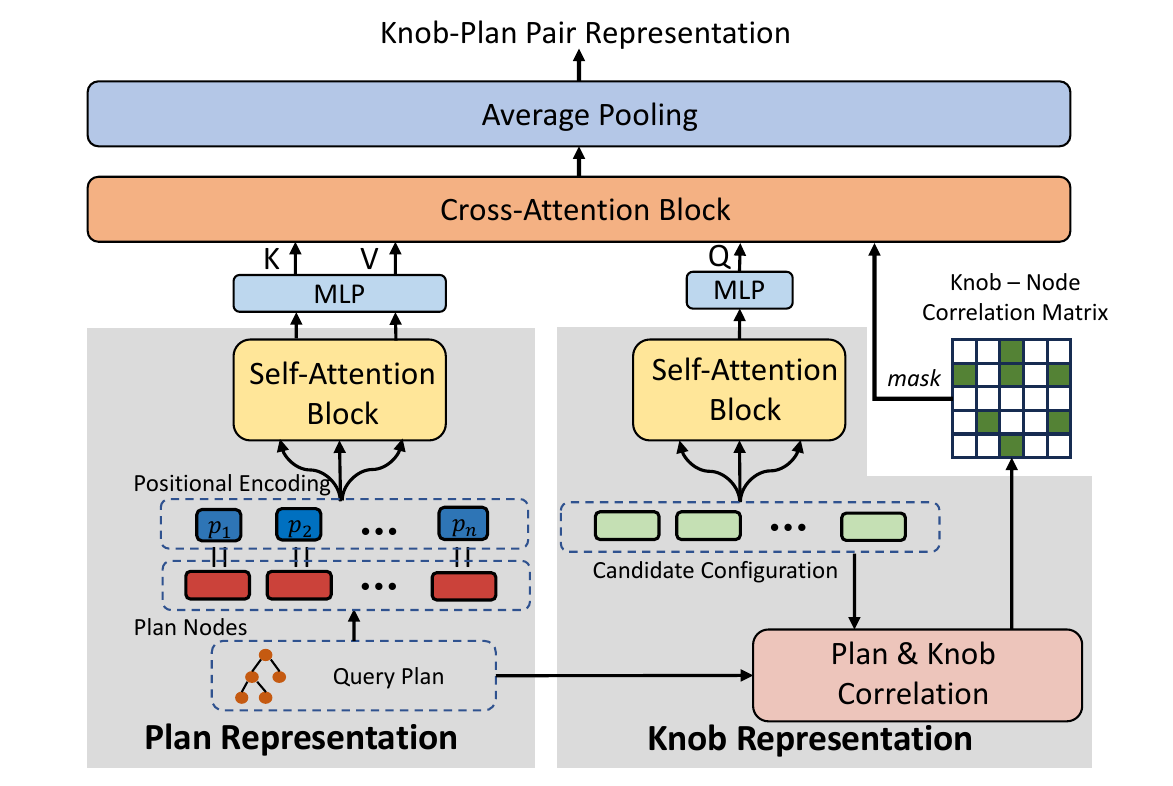}
  \vspace{-3em}
  \caption{The Knob-plan Joint Encoder.}
  \vspace{-2em}
  \label{fig: encoder}
\end{figure}

We now describe the architecture of the knob-plan encoder, shown in Fig.~\ref{fig: encoder}, which consists of three attention blocks.
Specifically, two self-attention blocks are used to separately encode the query plan and candidate configurations, generating vector representations of plan nodes and knobs. Both outputs are mapped to a uniform dimensionality using Multilayer Perceptrons (MLPs).
Then, the cross-attention block above them jointly encodes the knobs and query plan based on their correlation, using outputs from the two self-attention blocks as inputs.
Finally, its output is consolidated into a single vector representation through the average pooling.

\subsubsection{Node Featurization}
A query plan node for analytical queries includes the following key elements as its features: the plan operator type, involved tables and columns, predicates, join conditions, aggregation functions, and cardinality/cost estimations.
To encode these elements, we use the one-hot encoding for the plan operator type, tables, columns, join conditions, and aggregation functions, 
due to their relatively static and finite domains under a specific analytical query engine and workload context.
Predicates are represented as triplets $\langle column, comparison\_operator, value \rangle$, where the comparison operators (e.g., $>$, $<$, $=$) define the conditions.
Each element of the triplet is individually encoded and concatenated to form the full predicate encoding.
Continuous variables, such as the values in predicates and cardinality/cost estimations, are normalized to the range $[0, 1]$.
\yuxing{To handle string-like queries, we employ a trie-based encoding scheme~\cite{wang2021face}, where column strings are organized in a prefix tree structure. In this structure, each node represents a prefix, and leaf nodes are assigned unique IDs along with associated ranges that correspond to matching strings. 
This structure enables the transformation of string-like queries into range queries for efficient encoding.}

All feature encodings are concatenated into a single vector to create the node encoding $n_i$ for each plan node $i$.
Note that if a node does not contain specific information, such as predicates or join conditions, zeros are padded to guarantee consistent dimensionality across all node representations.

\subsubsection{Hierarchical Spectral Positional Encoding} 
Given the encoding of each node, we employ the self-attention to encode the entire query plan.
To reduce the computation cost, we only compute the attention scores between the connected nodes~\cite{mo2023lemo}.
However, preserving information of the plan's structure and the relationships between nodes is challenging.
The standard attention mechanism, designed for sequential data, cannot be directly applied to tree-structured plans.
To this end, we introduce a {\uline{H}ierarchical \uline{S}pectral \uline{P}ositional \uline{E}ncoding} (\HSPE) scheme to incorporate positional information of plan nodes.
This scheme combines two components: hierarchical and spectral encoding.
The hierarchical encoding employs Breadth-First Search (BFS) to capture the level-based structure, while the spectral encoding uses Laplacian eigenvectors to capture the tree's global properties, such as connectivity patterns between large subtrees or major branches.
This positional encoding scheme effectively combines local neighborhood information with long-range dependencies across the entire graph.



Specifically, for each node $i$, we construct its \HSPE as $\mathbf{p}_i$ = [$\mathbf{p}_i^{\operatorname{BFS}}||\mathbf{p}_i^{\operatorname{Lap} }$].
The hierarchical encoding, $\mathbf{p}_i^{\operatorname{BFS}}$, represents the BFS distance from the root to node $i$, capturing its level information in the tree.
For the spectral encoding, we first compute the Laplacian matrix $L$ of the plan tree, which captures the global structure by reflecting the connectivity between nodes, both directly and indirectly. Next, we compute the eigenvectors and eigenvalues of $L$, where each eigenvector represents a specific connectivity pattern across all nodes, and the corresponding eigenvalue indicates the frequency of these patterns. We select the $k$ eigenvectors associated with the smallest non-zero eigenvalues $v_1$, $v_2$, ..., $v_k$, which capture the most significant global structure in the plan tree. 
The spectral encoding for each node $i$ is then represented as $\mathbf{p}_i^{\operatorname{Lap}}$= 
($v_{1i}$, $v_{2i}$, ...,$v_{ki}$), where $v_{ji}$ denotes the value of $j$-th eigenvector at the $i$-th node. 
Note, constructing the Laplacian matrix for a tree and computing its eigenvalues and eigenvectors is efficient, as it scales linearly with the number of plan nodes.
Finally, we concatenate each node's \HSPE with its original encoding as $n'_i=n_i||\mathbf{p}_i$ and use it as the input of the self-attention block. 
After applying the self-attention mechanism, we obtain a sequence of node embeddings, the length equal to the number of nodes in the query plan.

\subsection{Knob Representation}
The self-attention mechanism is also employed to encode the candidate knobs, which helps capture the inter-dependencies between knobs.
To construct a vector representation for each configuration as the input of the attention block, one straightforward choice is the one-hot encoding. However, this method could not preserve the specific knob value defined in the continuous spaces.

Instead, we employ a variant of one-hot encoding called \textit{weighted one-hot encoding}, where the value $1$ is replaced with the actual configuration value.
For instance, consider a set of candidate knobs $k_1$, $k_2$, and $k_3$ with normalized values $0.3$, $0.6$, and $0.9$.
Then, we can encode these values into representation vectors as $[0.3, 0, 0]$, $[0, 0.6, 0]$ and $[0, 0, 0.9]$. 
In cases where the number of knobs to be tuned is large, resulting in especially high-dimensional knob embeddings, an autoencoder~\cite{hinton2006reducing} can be employed to reduce the high-dimensional embeddings to a more manageable and lower-dimensional space.
Note, positional encodings are not applied in encoding knobs since there is no inherent ordering among them.
Similar to the plan representation, the output of the self-attention block is a sequence of knob embeddings, with the length equal to the number of knobs.

\subsection{Synergizing Plan and Knob}  \label{section:conf-plan}

To capture the influence of the knob on the query plan, the knob-plan encoder employs the cross-attention block to synergize the knob and query plan,  generating a joint encoding for the downstream predictor.
%
To apply the cross-attention, we treat the sequence of plan embedding as the source sequence and the sequence of knob embedding as the target sequence. 
The cross-attention $A^c_{ij}$ score represents the correlation between the configurable knob $k_i$ and query plan node $n_j$.
There are two reasons for introducing this kind of attention.
First, the goal of \AQETune's encoder is to generate sufficiently informative representations of the knobs, which enables a more accurate prediction of how different knob configurations affect the query performance. Second, the sequence of plan embeddings is usually longer, as the number of nodes in an analytical query plan often exceeds the number of knobs involved in the BO-based tuning process.
With this mechanism, it allows the shorter sequence to be interpreted in the context of the longer sequence with richer information~\cite{gheini2021crossattn}.




However, it is unnecessary to compute the attention score between every knob-plan pair, since each knob typically impacts only a limited number of query nodes.
For example, adjusting a knob that controls filter strategy will affect only the execution time of nodes containing filter-related operators.
To this end, we introduce a \textit{correlation matrix} $M$ that illustrates whether the specific configurations affect the execution runtime of each node.
This matrix is further used to mask attention scores, ensuring only relevant interactions are considered. 
Specifically, $M_{ij}$ is set to $0$ if knob $k_j$  has no impact on the node $n_i$, and 1 otherwise.
Given the cross-attention score matrix $A^c$, its masked formula is computed as:
\begin{equation}
	 A_{ij}^{c'} = M_{ij} \cdot {A}^{c}_{ij}. 
\end{equation}

To construct the correlation matrix, one could manually gather the knowledge from DBAs and determine which tuned knobs would affect a given plan node. However, this approach is labor-intensive and inefficient. To address this issue, we propose an automated SHAP-based method (see Section~\ref{section:warm_starter}) to automatically identify the impact of knobs on the plan nodes by the warm starter.

%% file: sections/np_tune.tex
\section{Dual-Task Predictor}
\label{section:np_tuner}
We propose a dual-task predictor that predicts both the reliability and performance of analytical queries, using the joint representation from the knob-plan encoder as input.
In this section, we first give a conceptual overview of the Neural Process, which is the base model of our predictor.
Then, we explore the modeling strategy for dual-task prediction.
Lastly, we present the details of our predictor structure.

\label{section:np}



\begin{figure*}[htbp]
    \centering \captionsetup[subfigure]{skip=-0.1em}
    \subfloat[\revise{Vanilla Neural Process}]{
        \includegraphics[height=0.2\textwidth]{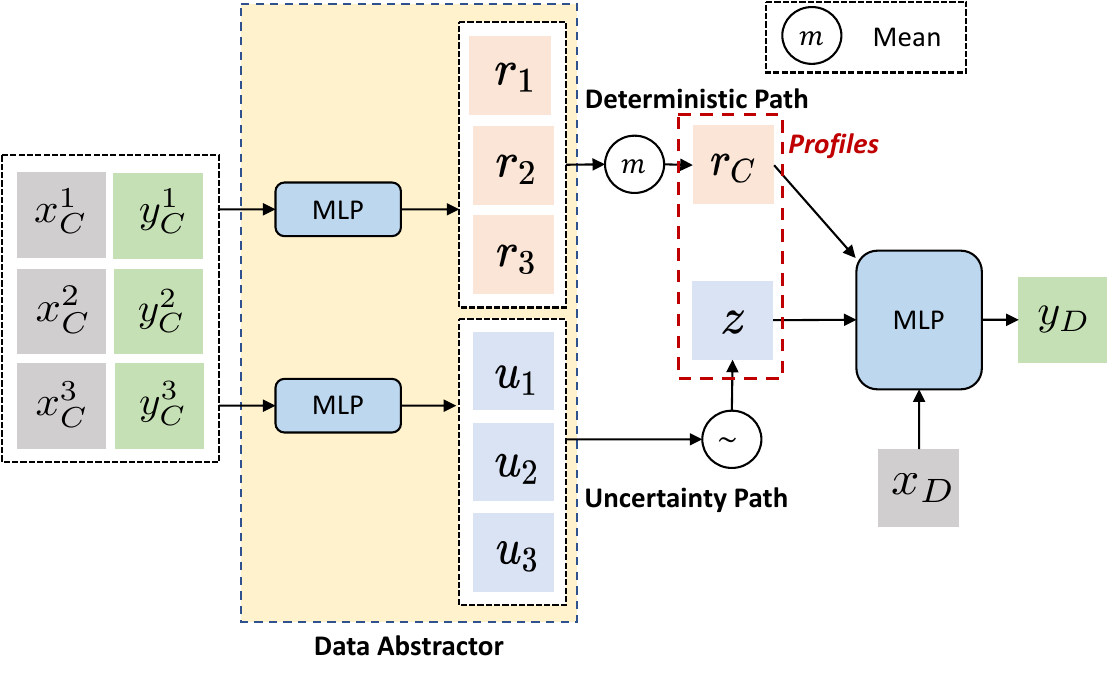}
        \label{fig:np}
    }
\hspace{0.3cm}
    \subfloat[\revise{Dual-Task Predictor}]{
        \includegraphics[height=0.2\textwidth]{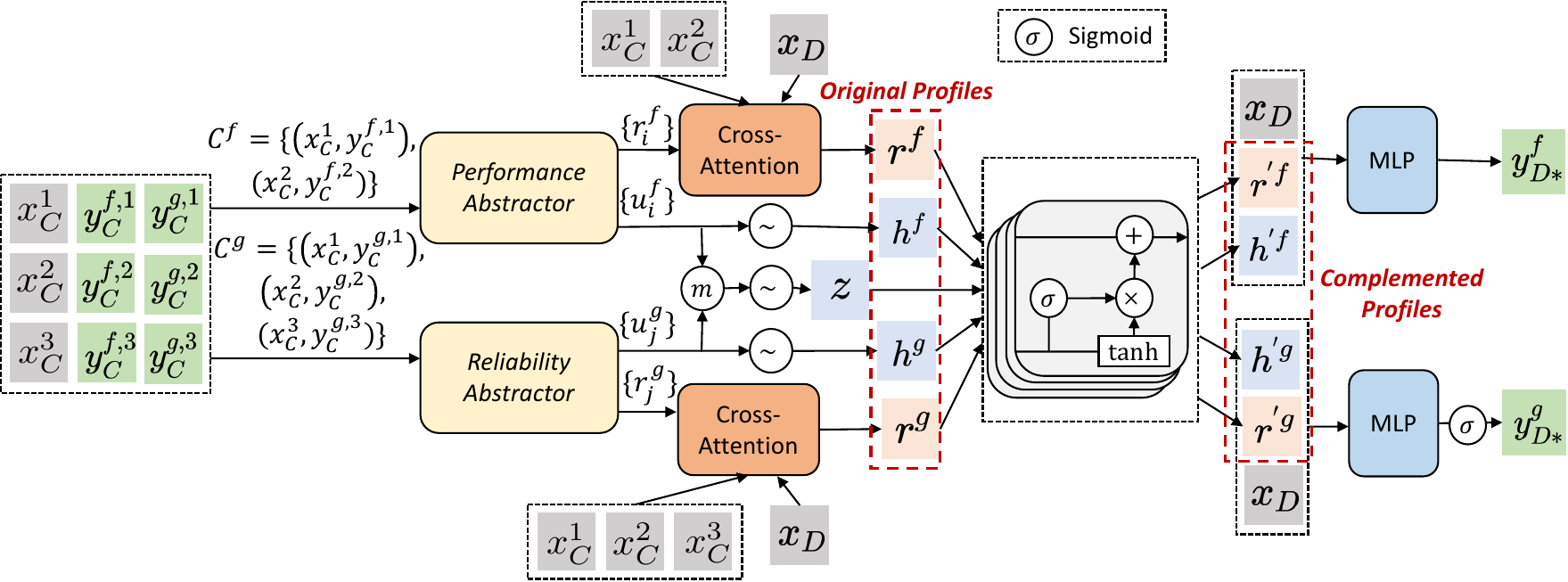}
        \label{fig:dual-task}
    }
    \vspace{-1.6em}
    \caption{Model Architecture.}
    \vspace{-1.8em}
    \label{fig:nn}
\end{figure*}

\subsection{\yuxing{Neural Process Overview}}


\yuxing{
Neural Processes (NPs) are a class of probabilistic models that blend the expressive power of neural networks with the probabilistic reasoning of stochastic processes (e.g., Gaussian Processes). 
The goal of an NP is to learn from a set of input-output pairs and subsequently predict the output for unseen input points.
%
}
%
The core structure of a vanilla NP (as shown in Fig.~\ref{fig:np}) revolves around two parallel paths.
Formally, given a set $C$ of context pairs $(x_C, y_C)$, NP defines a family of conditional distribution $p(y_D|x_D, C)$, where $(x_D, y_D)$  is a target pair in the set $D$.
To model the conditional distribution, NP utilizes a \textit{data abstractor} to independently encode context pairs 
using MLPs for two data paths:
\begin{itemize}[left=0em, topsep=0pt]
\item 
\textbf{Deterministic Path.}
\yuxing{This path encodes the input-output pairs from the context set into a global representation $r_C$, enabling the model to capture context-specific information.
By leveraging this representation, the model can identify patterns directly from the observed data, enhancing predictive accuracy.}
\item 
\textbf{Uncertainty Path.} 
\yuxing{This path incorporates a global latent variable $z$, shared across all context pairs, to encapsulate information not directly observable from the context set.
$z$ is commonly modeled via a variational distribution~\cite{hoffman2013stochastic}, which learns an approximate posterior distribution conditioned on the context set.}
\end{itemize}
We refer to both $r_C$ from the deterministic path and $z$ from the uncertainty path as \textit{profiles} of the context pairs.
The objective of NP is to maximize the likelihood of the target outputs $y_D$, conditioned on the context set $C$ and the corresponding target inputs $x_D$ from $D$.
Formally, the likelihood of $y_D$ is defined as:
\begin{equation}
	p(y_D \mid x_D, C) = \int p(y_D \mid x_D, {r_C}, {z}) q({z} \mid C)d{z}.
\end{equation}
where
$q$ represents a variational distribution from which the latent variables $z$ are inferred based on the context $C$.
%
For a given target $x_D$, the prediction of $y_D$ is generated using an MLP that takes $r_C$ and $z$ as inputs.
The parameters of an NP can be learned by maximizing the following Evidence Lower BOund (ELBO)~\cite{garnelo2018nps, kingma2013vlbo}:
\begin{equation}
\begin{array}{l}
\log p\left(y_{D} \mid x_D, C\right)
\geq \\ \mathbb{E}_{q(z \mid D)}\left[\log p\left(y_D \mid x_D, {r_C},{z}\right)\right]  \quad \quad \\ -\operatorname{KL}\left(q(z \mid D) \| q(z \mid C)\right),
\end{array}
\end{equation}
where the Kullback-Leibler (KL) term acts as a regularizer, ensuring that the distribution of the context data remains close to that of the target data.




\subsection{\yuxing{Joint Modeling of Performance \& Reliability}}


We model the task of the dual-task prediction with joint output as: $x \xrightarrow{} y^{f} \bigcup y^{g}$, where $x$ is the joint encoding generated by the knob-plan encoder, $y^f$ and $y^g$ represent the two prediction objectives:  performance and reliability. 
Drawing inspiration from the Neural Process (NP), we introduce two kinds of latent variables for dual-task prediction: cross-task and intra-task.

\yuxing{The cross-task latent variable, denoted as $z$, captures global uncertainty that is shared between query performance and reliability prediction tasks. It allows the model to recognize patterns that affect both tasks simultaneously.
For example, increased operator parallelism influences both the query execution time and reliability. By capturing these shared factors, $z$ serves as a bridge for sharing information across both prediction tasks, enabling them to learn from each other and improve predictive accuracy.}

\yuxing{Intra-task latent variables, denoted as $h^f$ for query performance and $h^g$ for reliability, are introduced to capture task-specific uncertainty unique to each prediction task. 
Unlike the cross-task variable $z$, which models shared global factors, $h^f$ and $h^g$ captures the unique features of each task. For example, $h^f$ focus on the factors that affect performance, such as data distribution patterns and hardware configurations, while $h^g$ focuses on the factors related to reliability, such as memory thresholds and CPU usage constraints that may trigger query failures. This separation allows the model to better understand and predict the unique behaviors of each task.
}


Formally, given an observation context set $C$ containing performance and reliability context pairs  
$(x_C, y^f_C)$ and $(x_C, y^g_C)$, and a target set $D$ of pairs  $(x_D, y^f_D)$ and $(x_D, y^g_D)$,
the prediction likelihood over two tasks can be expressed as:


\begin{equation}
\vspace{-1.6mm}
\begin{array}{c}
p(y^{f}_{D}, y^{g}_{D} \mid x_{D}, C) = \\ 
\quad \quad \prod_{t \in \{ f, g \}} \int \int p(y_{D}^{t} \mid h^{t}, x_{C}^{t}) p(h^{t} \mid z)  p(z \mid C) d h^{t} d z ,
\end{array}
\label{equation:likelihood}
\end{equation}

\noindent \textbf{Discussion.}
Before delving into our dual-task predictor structure, 
we first clarify the rationale behind this modeling strategy.
A straightforward approach is to treat the two prediction tasks independently and utilize two separate NPs to perform each prediction task.
However, this assumption is improper since reliability and performance are inherently correlated.
For instance, a knob adjustment aiming to improve query performance might negatively affect reliability by increasing resource usage.

Another intuitive approach is to combine the outputs of both tasks into a single output space by concatenating the two objectives as: $y^{f} \| y^{g}$.
This could be handled by a single NP model.
However, this formulation also has its limitations.
Concatenating the outputs implies that both observed objectives are equally informative, which is not always true. In cases of query failure, the observed query performance is usually not informative, as the query is only partially executed.
Adding such observations to the training dataset would worsen the model performance.
Therefore, the concatenated output method is not practical.

\subsection{Predictor Structure}


Figure~\ref{fig:dual-task} presents the architecture of our dual-task predictor.
The model employs two data abstractors, each with deterministic and uncertainty paths, to independently encode the performance and reliability context data $C^f$ and $C^g$.
To enhance the deterministic paths, the cross-attention mechanism is used in each task, enabling the model to focus on context data relevant to the specific target. 
For the uncertainty path, intra-task variables $h^f$ and $h^g$ help capture task-specific uncertainty, while the cross-task variable $z$ is introduced to capture the shared information between the two tasks.
In addition, since query performance and reliability are inherently correlated, we utilize a gating mechanism to complement the profiles in the paths of the same type. This approach allows information sharing between the two tasks, which could further help improve both of their prediction accuracy.
These complemented profiles are subsequently fed into final MLPs to generate the prediction outputs.

Next, we provide more technical details of the predictor structure, using an observation dataset comprising three instances and a target dataset with one instance $(x_D, y^f_D, y^g_D)$ as the example.
We assume that the last instance in the observation dataset is a failed query with $y_C^{g,3}=1$. 
Note, in practice, there usually exist multiple instances in the target dataset.


\noindent\textbf{Two Independent Data Abstractors.} 
We first remove the performance data from the observation instances that contain query failures,
as such data does not provide meaningful information. 
After this preprocessing step, we split the example observation data into two sets: $C^f=\{(x^1_C, y^{f,1}_C), (x^2_C, y^{f,2}_C)\}$ for performance and $C^g=\{(x^1_C, y^{g,1}_C), (x^2_C, y^{g,2}_C), (x^3_C, y^{g,3}_C)\}$ for reliability.
Then, we utilize two \textit{data abstractors} to encode the observation data of performance and reliability, respectively.
Similar to the vanilla NP in Fig.~\ref{fig:np}, each data abstractor consists of two MLPs.
That is, for each context pair $(x^i_C, y^{f,i}_C)$ in $C^f$ and $(x^j_C, y^{g,j}_C)$ in $C^g$, we would obtain their representations $r^f_i, u^f_i $  and $r^g_j, u^g_j$ through the \textit{data abstractors},
which are subsequently used to process the deterministic and uncertainty paths for each prediction task.

 


\noindent \textbf{Cross-attention Aggregator.} 
For the deterministic paths of both prediction tasks, we replace the mean aggregator used in the vanilla NP with the cross-attention mechanism.
Unlike the mean aggregator, the cross-attention allows each target data item $x_D$ to selectively focus on those context data items that are most relevant to predicting its performance and reliability output $y_D^f$ and $y_D^g$.
We do not use cross-attention in the uncertainty path because this path aims to capture the overall uncertainty that affects all the target predictions collectively rather than each prediction individually~\cite{kim2019attentivenp}.
 
\noindent \textbf{Complemented Profiles.}
%
\yuxing{Profiles from the same type of path in the two tasks could provide complementary information, which further helps improve the accuracy of each task's prediction.
To achieve this, we utilize a gating mechanism to facilitate information sharing across the paths.
While the context representations ${r}^f$ and ${r}^g$ are generated from the two deterministic paths independently, they often contain useful information for each other.
The gating mechanism allows $r^f$ to ``borrow'' useful features from $r^g$ while filtering out irrelevant parts.
For example, if $r^g$ identifies high memory consumption, this information may also be useful for predicting slow query execution.
By controlling how much of $r^g$ is incorporated into $r^f$, the model avoids overfitting to noise while still benefiting from useful shared features.
This information-sharing strategy is not limited to context representations ($r^f$, $r^g$) but is also applied to latent variables ($h^f$, $h^g$, and $z$), ensuring
only relevant information flows between them across all model paths.}


Specifically,  we define four profile pairs: ($r^f$, $r^g$), ($r^g$, $r^f$), ($h^f$, $z$) and ($h^g$, $z$). 
To enable dynamic information sharing between each pair, the gating mechanism adjusts the influence of the second profile on the first as follows: 
\begin{equation}
\begin{array}{c}
p_o'= p_o + \tanh (W_1 p_c + b_1) \times \sigma (W_2 p_c + b_2),
\end{array}
\vspace{-2mm}
\end{equation}
where $\sigma$ denotes the sigmoid function, $W_1$, $W_2$ are weight matrices, and $b_1$ and $b_2$ are bias vectors. We refer to $p_o'$ as the complemented $p_o$.
Then, after applying the gating mechanism on the four profile pairs mentioned above, we obtained complemented representations $r'^f$, $r'^g$, and complemented latent variables $h'^f$, $h'^g$.
%

%
%

\noindent\textbf{Prediction Generator.} Finally, two generators are used to process the four complemented profiles, producing prediction outputs for both tasks.
Specifically, the performance prediction is handled as a regression task, while the reliability prediction is treated as a binary classification task. 
Each task's complemented profiles, along with the target data item $x_D$, are fed into two separate MLPs.
Note, the reliability prediction task uses an additional sigmoid activation function to map the MLP's output into the range $[0, 1]$.


\subsection{Model Training}
We then explain how to end-to-end train our \AQETune model, including the knob-plan encoder and dual-task predictor.
Specifically, the dataset is randomly divided into an observation set and a target set, each comprising 50\% of the data.
The model takes the observation set $C$ and $x_D$ from the target set as input, while $y_D$ from the target set serves as the ground truth. 
As directly optimizing the likelihood in Eq.\ref{equation:likelihood} is computationally prohibitive, we approximate it using the variational lower bound~\cite{shen2024episodic}:

\begin{equation}
\begin{array}{l}
\log p_{\theta}\left(y^{f}_{D}, y^{g}_{D} \mid x_{D}, C\right) \\
 \geq \mathbb{E}_{q_{}(z \mid D)}\bigg [\sum_{t\in \{f,g\}} \Big [\mathbb{E}_{q_{}\left(h^{t} \mid z, D^{t}\right)}\big[\log p_{\theta}\big(y_{D}^{t} \mid x_{D}, {h}^{t}, {r}^t\big)\big] \\ \quad \quad -  \operatorname{KL} \left(q_{}\left({h}^{t} \mid z, D^{t}\right) \| q_{}\left({h}^{t} \mid z, C^{t}\right) \right) \Big ] \bigg] \\ \quad \quad-\operatorname{KL}\left(q_{}(z \mid D) \| q_{}(z \mid C)\right).
\end{array}
\end{equation}
This equation involves two levels of expectations: first over the intra-task variables $h^t$, and then over the cross-task variable $z$.
The KL divergence terms act as regularizers, ensuring that the learned distributions for both intra-task variables and the cross-task variable remain close to their observation distributions, respectively.

\yuxing{
\subsection{Generalization To New Datasets}
Real-world analytical query engines handle diverse datasets with varying schemas and query workloads, making it crucial for tuning systems to generalize to unseen datasets.
\AQETune achieves this generalization in two ways. 
First, the knob-plan encoder encodes the query plans with the self-attention mechanism and positional encoding \HSPE.
Since query plans reflect SQL operations rather than specific data content, they naturally generalize to new workloads. 
Besides, the encoder employed the cross-attention mechanism to establish correlations between query plans and knobs, enabling \AQETune to generalize to unseen tuning cases by learning the impact of knobs on query execution.
Second, the dual-task prediction is built on NPs to predict both query performance and reliability. 
In essence, NPs learn a distribution over possible functions based on observed data rather than a fixed input-output mapping~\cite{garnelo2018nps}. This approach allows the predictor to adapt to unseen datasets by selecting function mappings that best adapt to new observations, enhancing \AQETune's generalization capability.
}

%% file: sections/pso_tune.tex
\section{Warm Starter} \label{section:warm_starter}

In this section, we introduce the warm starter, which generates high-quality initial observation data for BO-based tuning and pre-calculates the correlation between knobs and query plans.

\subsection{Generating Initial Samples}
At the initial stage of BO-based tuning for each target query, high-quality observation data is critical as it provides the first insights for \AQETune's tuning objectives.
To generate high-quality observation data, we propose sampling points in the knob space $\Theta$ that should meet two key criteria: \textit{diversity} across the search space and \textit{efficiency} to explore potentially good sample regions. 
The diversity criterion ensures that the initial samples cover the search space as much as possible, enabling the surrogate model to capture the overall shape of the objective function.
The efficiency criterion involves quickly identifying regions with potentially good performance, typically around previously observed promising samples.




However, existing sampling methods could not meet the two criteria simultaneously. Specifically, most methods~\cite{zhang2021restune, van2017ottertune} rely on random sampling to collect initial points. Although random sampling meets the diversity criteria, it does not efficiently explore promising regions.
Hunter~\cite{cai2022hunter} tries to address this issue by using a Genetic Algorithm to target promising regions during sampling.
However, its sequential sampling process requires evaluating one point before the next can be sampled, limiting efficiency and reducing the number of samples explored within a given time period.


To address this issue, \AQETune leverages Particle Swarm Optimization (PSO)~\cite{shami2022particle} for initial sample collection.
PSO is an optimization method that iteratively tries to improve candidate samples based on a given quality measure.
It has the strength of fast convergence without requiring prior knowledge.
In PSO, a group of candidate sample points, known as \textit{particles}, move through the search space by adjusting their positions based on their velocities.
The velocity determines both the direction and speed of the particle's movement.
Each particle acts as an independent agent, which maintains its best-known position and could also see the best-known position among all particles.
The movements of particles are influenced by this information, guiding them to promising regions. In this way, the particles could produce promising samples during the optimization process. Since each particle is evaluated independently, PSO is particularly well-suited for parallelization.

PSO could perfectly meet the two key criteria in our problem. The initial positions of particles are generated randomly to ensure \textit{diversity}.
For the \textit{efficiency} of sampling, PSO excels in parallel search, allowing particles to explore promising regions independently.
Once a particle completes its evaluation, it immediately updates its next sampling position based on global and local optima, without waiting for others.
This ensures fast and efficient exploration throughout the search process.

\begin{algorithm}[t]
\small
  \caption{\yuxing{PSO-based Sampling.}}
  \label{alg: heuristic-tuning}
   \setlength{\abovecaptionskip}{0.cm}
  \SetAlgoLined
  \KwIn{
  A target query $q$;
  Number of required samples $M$; 
  Number of particles $P$;
  }
  \KwOut{Configuration Samples $\mathcal{S}$.}
  
\For{each particle $i = 1$ to $P$}
{
	randomly sample a position $s_i$ and a velocity $v_i$\;
}

\While {$|\mathcal{S}| < M$} {
  \For{each particle $i = 1$ to $P$}
  {

		($f_i$, $g_i$) = EvaluteOnEngine($q$, $s_i$)\;
        $\mathcal{S} \gets \mathcal{S} \cup (s_i, f_i, g_i) $ \;

        
%
        
        
        \If {$g_i = 0$}
        {
        	Update particle specific best position $LB_i$ if $f_i$ is the best performance in the particle $i$\;
            Update global best position $GB$ if $f_i$ is the best performance so far\;
            
            $v_i = \mu \cdot v_i + c_1 \cdot r_1 \cdot (LB_i - x_i) + c_2 \cdot r_2 \cdot (GB - x_i)$ \;
            $s_i = s_i + v_i$ \;
        }
        \Else
        {
            randomly resample $s_i$ and $v_i$ \;
        }
        
        
    }
    
  }
  return $\mathcal{S}$ \;
\end{algorithm}



The procedure of PSO-based sampling for each target query is outlined in Algorithm~\ref{alg: heuristic-tuning}. The procedure has only one hyperparameter, which is the number of particles.
We start by randomly initializing the position of the configuration point and velocity for each particle (Lines 1$\sim$3).
Next, we evaluate each configuration sample with the analytical engine to obtain the query's performance and reliability (Lines 5$\sim$7). If the sample is reliable, which means that the query executes successfully under the sampled configuration, then we update the best position for this single particle and the global best position for all the particles (Lines 9$\sim$10). 
Here, the particle-specific best position refers to the optimal point found by an individual particle, while the global best is the optimal point found by all particles. 
Next, we adjust the position and velocity for the particle's next iteration (Lines 11$\sim$12). 
Note that the position update equations balance exploration and exploitation from particle-specific and global best positions in the search space. Here, $c_1$ and $c_2$ are local and global acceleration coefficients, which are used to determine the degree of influence that different kinds of best positions would have on the next move. Besides, $\mu$ determines how much of the previous velocity is retained, while $r_1$ and $r_2$ are random values between $[0,1]$ to add variability into the search process.
Finally, if the sample is unreliable, we resample the position and velocity  to continue the process (Line 15).



\subsection{Identifying Knob and Plan Node Correlation}
Based on the high-quality samples collected by the warm starter, we propose to identify a coarse correlation between knobs and plan nodes by calculating each knob's contribution to the execution time of different plan nodes (e.g., filter, join). 
This is essential for building the correlation matrix used by the cross-attention mechanism within the knob-plan encoder (see Section~\ref{section:conf-plan}).
To this end, we evaluate the target queries with the sampled configurations using the ``explain analyze'' statement, which provides the execution time $t$ of each plan node. 
For each node type $nt$, we organize the evaluated results as a triplet ($nt$, $s$, $t$). Here, $s$ is a vector with each element $s_i$ representing a normalized knob value.

Next, we propose to apply the SHAP~\cite{lundberg2017shap} method to identify the importance of each knob by computing its contribution to the prediction of node's execution time. This is because SHAP has been proven to be robust in model explanations~\cite{zhang2021treeshap}. 
In our scenario, each knob value $s_i$ is treated as a feature to predict the node's execution time. 
Then, we utilize SHAP to calculate the knob's contribution values for each node.
If the contribution value is non-zero, it indicates that the knob has influences on the execution behavior of the specific node type, and they are considered as correlated. Otherwise, they are not correlated.




\vspace{-2mm}
\subsection{\yuxing{Computational Complexity of \AQETune}}
\yuxing{The computational complexity of \AQETune is divided into three main components.
For the knob-plan encoder, given a query with $N$ nodes in its query plan and $|K|$ tunable knobs, the complexity is $\mathcal{O}(|K|^2 + N \cdot |K|)$.
This arises from the self-attention mechanism applied to the knobs and the cross-attention mechanism between the query plan nodes and knobs.
For the dual-task predictor, the primary computational cost comes from calculating the cross-attention scores between the observation and target datasets. 
The resulting complexity is $\mathcal{O}(|C| \cdot |D|)$, where $|C|$ and $|D|$ denote the sizes of the observation and target datasets, respectively.
For the PSO-based warm starter, the complexity is dominated by the particle update step.
Assuming each evaluation on the analytical engine takes $T_{eval}$ time and the failure probability is $\lambda$, the 
total
complexity is $\mathcal{O}(\frac{M \cdot T_{eval}}{1-\lambda})$, where $M$ is the number of required samples.
}



%% file: sections/evaluation.tex
\section{Experimental Evaluation}

\noindent\textbf{Outline.} In this section, we first describe the experimental setup, followed by an evaluation of \AQETune's tuning efficiency across two analytical engines through an end-to-end tuning process. Lastly, we analyze the impact of our proposed three key components.



\noindent\textbf{Baselines.}
We evaluate \AQETune against a range of representative ML-based tuning systems, including:


  ResTune~\cite{zhang2021restune}: A BO-based system-level tuning method that considers the resource constraints during the tuning process. 
To reduce query failures, we add constraints on CPU utilization and memory usage based on the DBA experience. 
\yuxing{We used Latin Hypercube Sampling (LHS)~\cite{van2017ottertune} to collect initial tuning samples.}

 Hunter~\cite{cai2022hunter}: An RL-based system-level tuning method that applies the Genetic Algorithm for initial exploration to handle the cold-start issue. Besides, it utilizes DDPG as its neural network architecture.

 QTune$^{Q}$~\cite{li2019qtune}: An RL-based tuning method that optimizes knob configurations at different-grained levels.
\yuxing{We choose its query-level version and use random sampling to collect initial tuning data.}

\yuxing{Bao~\cite{marcus2021bao}: A query-level hint selection method that converts the query plans into single feature vectors with tree convolutional neural network~\cite{mou2016convolutional} and pooling technique. 
To support knob tuning, we concatenated the candidate configurations with feature vectors to make predictions and used LHS to generate initial tuning samples.}

\yuxing{QueryFormer~\cite{zhao2022queryformer}: A Transformer-based model for query plan representation that supports query-level optimization. Similarly to Bao, we enabled QueryFormer to support knob tuning by concatenating plans' feature vectors with candidate configurations and used LHS to collect initial tuning samples.}




\noindent{\textbf{Workloads.}
We evaluate \AQETune using two synthetic benchmarks and two real-world workloads:
1) TPC-DS: An OLAP benchmark consisting of 99 queries, used to assess decision support system performance;
2) TPC-H: Another OLAP benchmark with 22 queries, simulating a business-oriented and ad-hoc query workload for decision support system evaluation;
3) STATS-Analytical: Derived from the STATS dataset~\cite{cardestbench}, originally designed for cardinality estimation with complex join schemas. We extend this workload with additional queries to meet practical analytical needs, such as calculating average post scores or the number of comments per post by year;
4) JOB-Extend: Based on IMDB~\cite{leis2015imdb}, originally used to test query optimizer performance. We extend the JOB workloads~\cite{leis2018query} to include more analytical tasks.
The data size for TPC-DS and TPC-H is set to 500 GB. Since the original  IMDB and STATS datasets are relatively small, we uniformly scale them to 100 GB.


\noindent\textbf{Environment.}
The experiments involve two analytical query engines, ByteHouse and Presto, each deployed in the pseudo-distributed mode with three instances.
Both engines run on a server with an Intel(R) Xeon(R) Platinum 8336C CPU @ 2.30GHz and 1TB of available memory.
Based on DBA expertise, 25 query-level knobs are selected for tuning in ByteHouse and 20 in Presto.
These knobs include the control of memory allocation, degree of parallelism, execution options for query operators in the plan, etc.

\noindent\textbf{Settings.}
We implement \AQETune using Python 3.8 and PyTorch 1.13.
\yuxing{To ensure a fair comparison, all experiments were conducted without any prior tuning data, and the tuning process for each workload, consisting of multiple queries, was constrained to a maximum duration of six hours.}
The execution time of failed queries is set as $100$ seconds to avoid demonstrating incorrect ``good'' query performance in the experimental results.

\yuxing{Once sufficient tuning data is collected, the query performance and reliability datasets are randomly split into training and test sets in a 7:3 ratio to train the knob-plan encoder.
The tuning dataset is constructed by pairing query plans with their corresponding knob configurations, which are used as inputs for prediction.
}
Each knob-plan pair in the encoder is represented with 32 dimensions, and we select 10 eigenvectors ($k = 10$) for the hierarchical spectral encoding of the plan.
For the warm starter, the number of particles is set to $3$, with coefficients  $\mu$, $c_1$, and $c_2$ set to $0.5$, $2$, and $2$, respectively.

\noindent\yuxing{\textbf{Metrics.}
For query performance, we analyze the average and 95th percentile (P95) of query latencies for a given workload. This analysis captures the effects of query-level tuning on overall processing and tail latency, which are critical for meeting the Service Level Objectives (SLOs) of database services~\cite{xu2024bouncer}.
For reliability, we track the number of query failures throughout the tuning process.}

\begin{figure*}
    \centering
    \captionsetup[subfigure]{skip=-0.3em} 
    \subfloat[JOB-Extend (ByteHouse)]{
    \label{fig:e2e_job}
    {\includegraphics[width=0.33\textwidth]{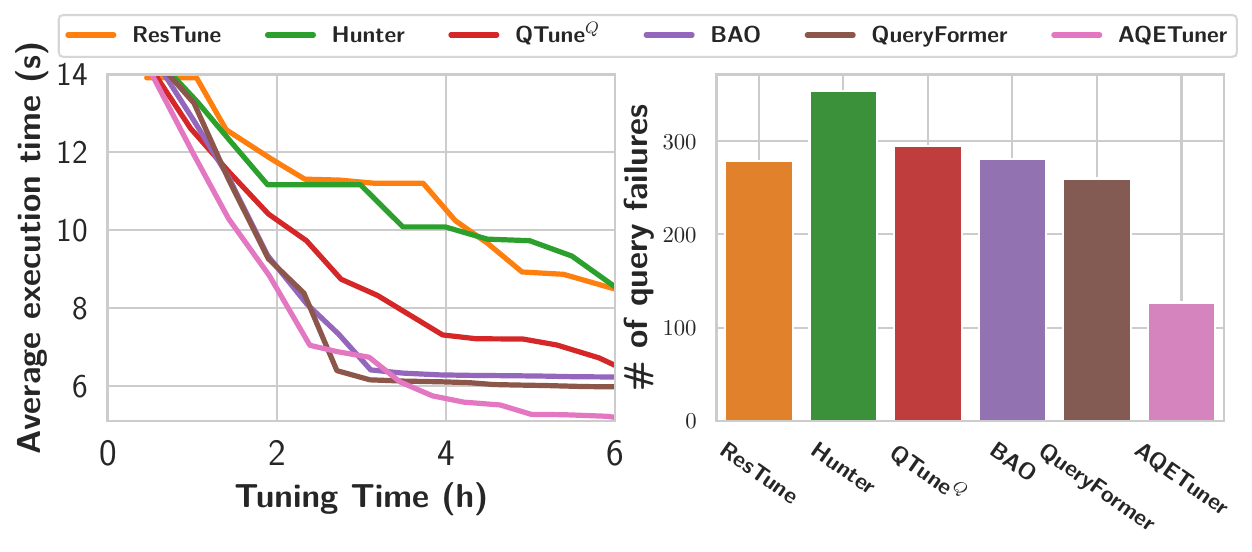}}}\hspace{-2.5mm}
    \subfloat[STATS-Analytical (ByteHouse)]{
    \label{fig:e2e_stats}
    {\includegraphics[width=0.33\textwidth]{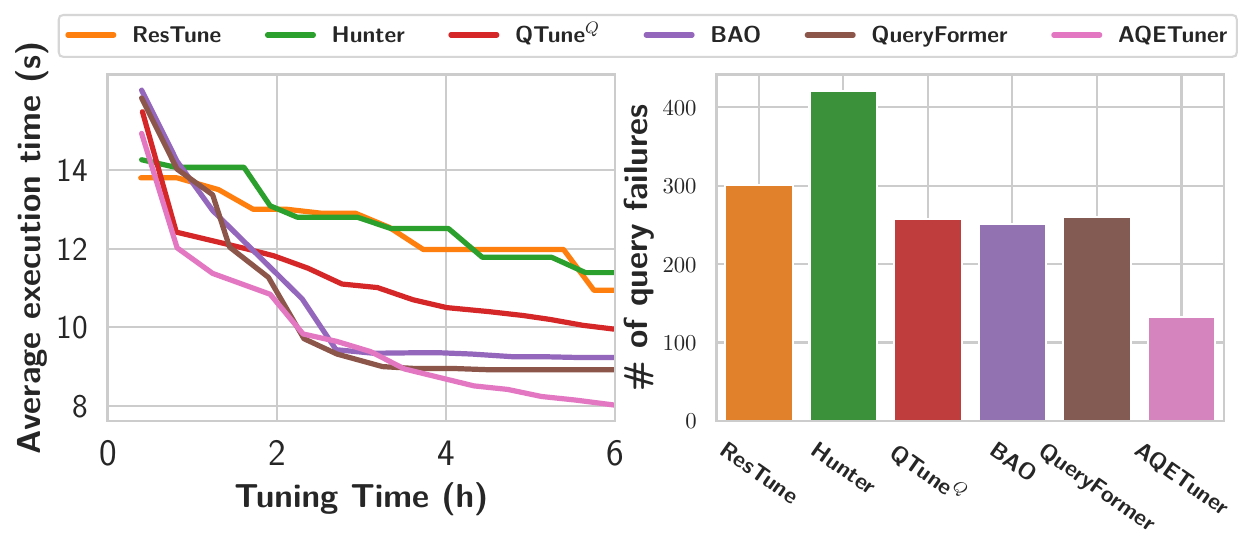}}} \hspace{-2.5mm}
    \subfloat[TPC-DS (ByteHouse)]{
    \label{fig:e2e_tpcds}
    {\includegraphics[width=0.33\textwidth]{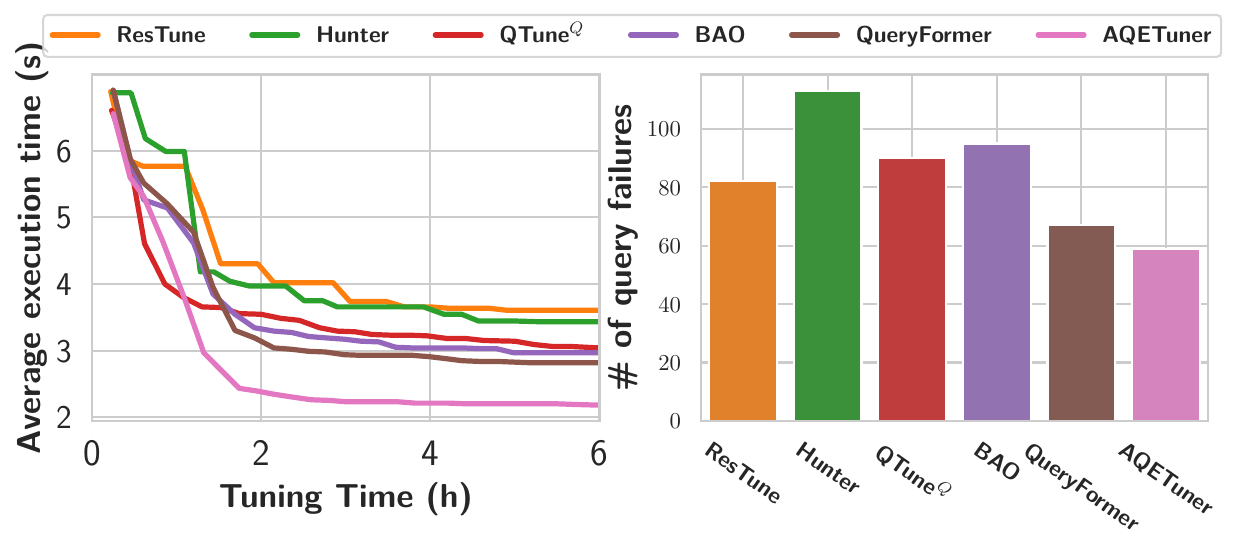}}} \\
    \subfloat[TPC-H (ByteHouse)]{
    \label{fig:e2e_tpch}
    {\includegraphics[width=0.33\textwidth]{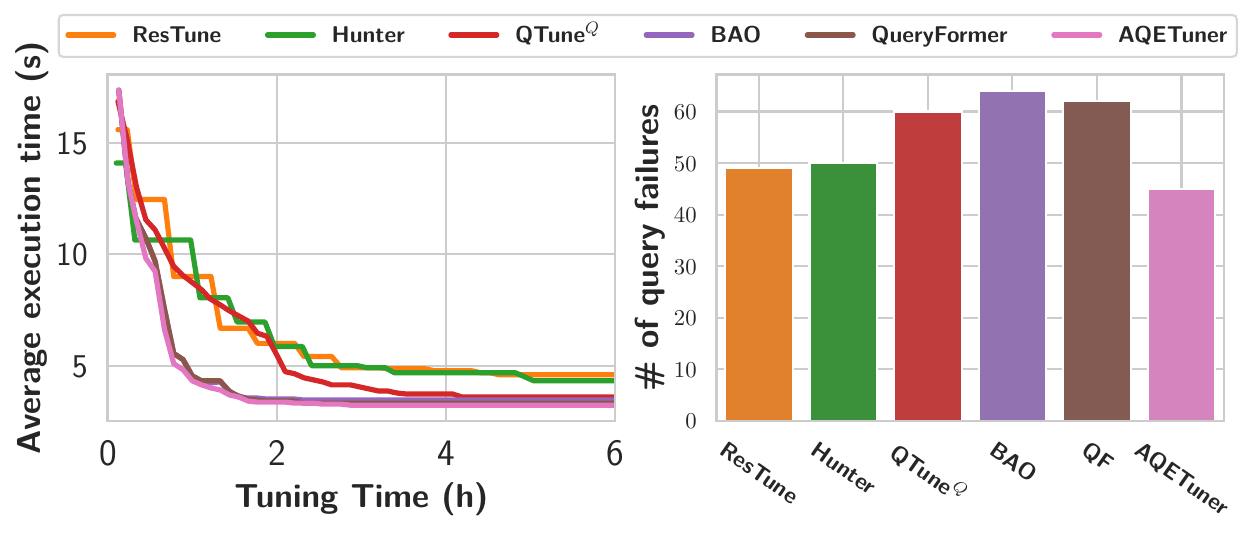}}}  \hspace{-3mm}
    \subfloat[TPC-DS (Presto)]{
    \label{fig:e2e_presto}
    {\includegraphics[width=0.33\textwidth]{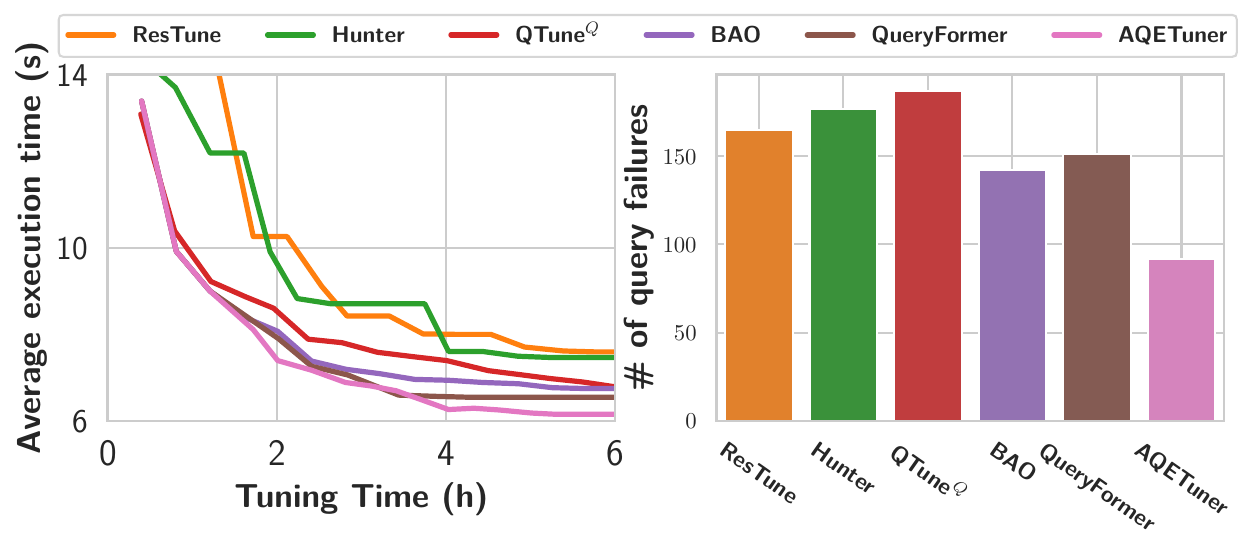}}} \hspace{-3mm}
    \subfloat[JOB-Extend P95 Latency (ByteHouse)]{
    \label{fig:e2e_tail_latency}
    \makebox[0.33\textwidth]{
    {\includegraphics[width=0.32\textwidth]{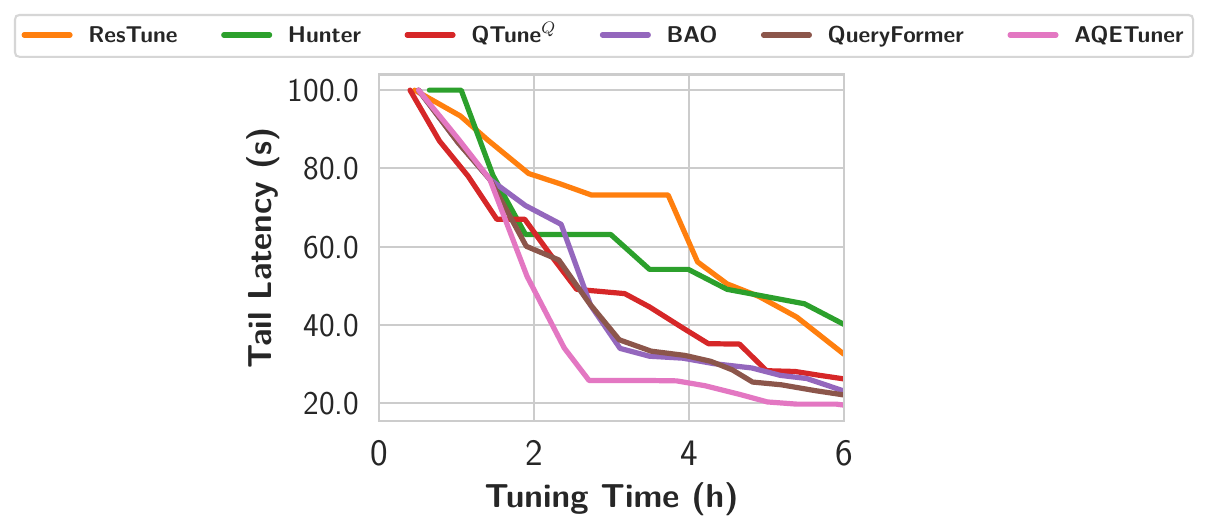}}}} 
    
\vspace{-1em}
\caption{Optimal execution time and the total number of query failures during tuning processes.}
\vspace{-1em}
\label{fig: e2e_experiments}
\end{figure*}

\vspace{-2mm}
\subsection{Efficiency Comparison}


\noindent\textbf{ByteHouse.}
Figure~\ref{fig:e2e_job}-\ref{fig:e2e_tail_latency} report the optimal average execution time achieved during the tuning process and the total number of query failures observed for different methods across various workloads in ByteHouse.
Overall, \AQETune outperforms all the baseline methods across all four workloads.
Notably, \AQETune achieves up to a 37.5\% improvement in optimal execution time compared to system-level methods and \revise{up to a  28.3\% improvement over three query-level competitors.} Regarding query failures, \AQETune reduces the number by up to \revise{51.2\%} compared to the next-best method. The details are as follows:

1) System-level tuning methods show similar but generally worse performance than query-level methods. 
Query-level tuning excels by collecting more observations and using query plans instead of SQL text for modeling. 
In system-level tuning, the average execution time of all queries is considered a single observation per BO cycle, resulting in a limited number of observations throughout the tuning process. 
In contrast, query-level tuning treats the execution time of each individual query as an observation, enabling more data collection in one cycle.
Furthermore, query-level tuning methods utilize query plans, which capture more detailed execution information than system-level approaches. This enhances tuning efficiency and improves the chances of finding optimal configurations. 
As a result, \AQETune achieves optimal configurations that demonstrate up to 35.5\%, 29.7\%, 37.5\%, and 24.1\% lower average latency than the system-level methods under the JOB-Extend, STATS-Extend, TPC-DS, and TPC-H workloads, respectively.

2) \yuxing{
Among the query-level methods, \AQETune surpasses the other three ones (i.e., QTune$^Q$, Bao, and QueryFormer) for two main reasons.
First, these three methods either fail to capture the structure of the query plan or ignore the impact of knobs on plan node execution, limiting their tuning ability.
Second, they neglect the cold-start issue or rely on random sampling, resulting in suboptimal exploration during the initial phase.
In contrast, \AQETune leverages high-quality initial data from the PSO-based sampling method to accelerate exploration in subsequent BO-based tuning.}
%
\yuxing{As a result, 
\AQETune achieves optimal configurations that demonstrate up to 20.4\%, 19.4\%, 28.3\%, and 10.2\% lower average latency than other query-level methods under the JOB-Extend, STATS-Extend, TPC-DS, and TPC-H workloads, respectively.
}

3) To further validate \AQETune's tuning efficiency in ByteHouse, we evaluate the optimal 95th percentile of query latency (P95) on the JOB-Extend workload during the tuning process, as shown in Fig.~\ref{fig:e2e_tail_latency}. The query-level tuning methods consistently outperform the system-level methods.
\AQETune has  40\% lower P95 latency compared to the best system-level method.  
This is because system-level tuning methods do not target individual query performance as tuning objectives and cannot effectively optimize slow queries.
\yuxing{
Notably, \AQETune can reduce P95 latency up to 15.9\% and 12.3\% compared to Bao and QueryFormer, respectively.
}

4) For the number of query failures, \AQETune achieves the lowest count across all cases, but its advantage varies by workload. 
Compared to the next-best method, \AQETune reduces query failures by \revise{51.2\%} and \revise{47.5\%} on the JOB-Extend and STATS-Analytical workloads, respectively.
On TPC-DS and TPC-H, \AQETune's advantage is smaller, reducing failed executions by only \revise{15.5\%} and 8\%, respectively. 
This is because queries in TPC-DS and TPC-H are less likely to experience execution failures than those in JOB-Extend and STATS-Analytical, as they involve fewer multi-table joins, and thus demand less memory and computation resources.

\noindent\textbf{Presto.} To demonstrate the generality of \AQETune's tuning capabilities, we also evaluate it on another widely used analytical engine, Presto, using the community version 0.288. Since Presto only supports cloud storage module, we store the dataset on the HDFS for subsequent querying.
We compare the tuning efficiency of different methods on Presto under TPC-DS workloads, which have relatively complex query plans.
Note, we fail to run the JOB-Extend and STATS-Analytical workloads on Presto due to the relatively poor performance of cloud storage.
For the TPC-H dataset, we observed results similar to those of ByteHouse.
The evaluation results are shown in Fig.~\ref{fig:e2e_presto}.
For query performance, the average execution time for all methods shows a similar trend, decreasing as tuning progresses.
The performance ranking is exactly the same as that observed in ByteHouse: \AQETune > \revise{QueryFormer > Bao} > QTune$^Q$ > ResTune $\approx$ Hunter.
Specifically, the optimal configurations found by \AQETune result in 20.1\%, 17.6\%, 9.4\%, \revise{8.8\% and 6.1\%} lower average latency compared to
ResTune, Hunter, and QTune$^Q$, Bao and QueryFormer, respectively. 
Regarding query failures, \AQETune also achieves the lowest number, reducing failed executions by  
44.3\%, 48.1\%, 50.8\%, \revise{42.2\% and 46.7\%} compared to 
ResTune, Hunter, QTune$^Q$, \revise{Bao and QueryFormer}.


\vspace{-2mm}
\yuxing{
\subsection{Study of \AQETune's Tuning Cost}
To evaluate the tuning cost of \AQETune, we conducted two experiments on ByteHouse with the TPC-DS workload.
First, we divided \AQETune into three phases: warm start, model training (for the knob-plan encoder and dual-task predictor), and BO-based tuning.
Over a 6-hour query-level tuning session, we measured the memory usage and computational time for each phase.
As shown in Fig.~\ref{fig:tuning_parts_analysis}, the BO-based tuning is the most time-intensive, consuming 74.5\% of the total time and 3.2 GB of memory due to its extensive Bayesian Optimization exploration. 
Model training, while requiring 20.2\% of the time, exhibits the highest memory usage at 9.4 GB to support the encoder and predictor training.
In comparison, the warm start phase requires only 5.3\% of the time and 88.3 MB of memory, demonstrating its efficiency in initializing the tuning process.

Next, we analyzed the impact of the number of plan nodes and tunable knob on the average recommendation time in the BO-based tuning phase, as shown in Fig.~\ref{fig:eval_plan_nodes_knobs}.
The results reveal that the recommendation time increases significantly with the number of plan nodes, particularly when exceeding 100 nodes, highlighting a notable scalability challenge for complex query plans.
Besides, the relatively limited number of tunable query-level knobs has a comparatively smaller impact on the recommendation time.
}

\begin{figure}[tb]
\centering 
\captionsetup[subfigure]{skip=-0.1em} 
\vspace{-0.5em}
\subfloat[Different Tuning Phases]{
    \label{fig:tuning_parts_analysis}
    {\includegraphics[width=0.47\linewidth]{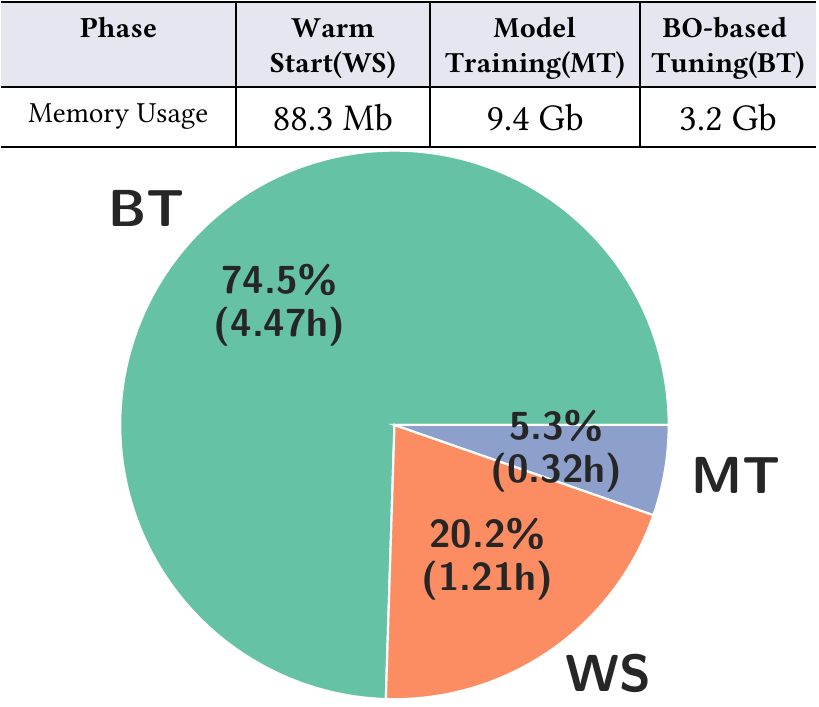}}}\hspace{1mm}
    \subfloat[{Effect of Plan Nodes and Knobs}]{
    \label{fig:eval_plan_nodes_knobs}    {\includegraphics[width=0.47\linewidth]{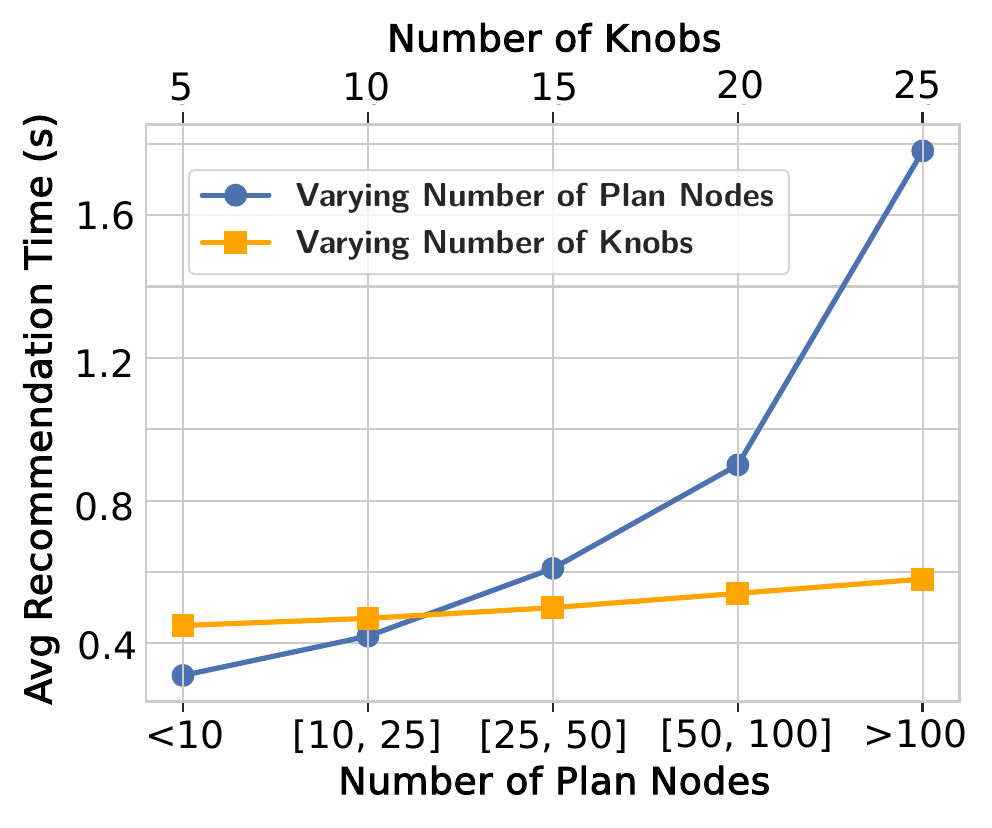}}}
  
  \label{fig: encoding_methods_effect}
  \vspace{-1em}
  \caption{\yuxing{Tuning Cost of \AQETune.}}
  \vspace{-1em}
\end{figure}

\vspace{-1em}
\subsection{Evaluation of Knob-Plan Encoder}
\AQETune's knob-plan encoder is the first to leverage the correlation between knobs and query plans to improve DBMS knob tuning. 
To validate its effectiveness, we compare it with two representative encoders: the Query2Vec encoder for query encoding ~\cite{li2019qtune} and the QEP2Vec encoder for plan encoding~\cite{henderson2022blutune}.
The Query2Vec encoder converts SQL texts into feature vectors by encoding key elements (e.g., query type, involved tables, and operations).
The QEP2Vec encoder treats query plans as documents and sub-plans as words, applying word2vec~\cite{mikolov2013word2vec} to learn the plan embeddings. 


\begin{figure}[tb]
\centering \captionsetup[subfigure]{skip=-0.3em} 
\subfloat[{Query Performance Accuracy}]{
    \label{fig:encoding_method_box_data}
    {\includegraphics[width=0.47\linewidth]{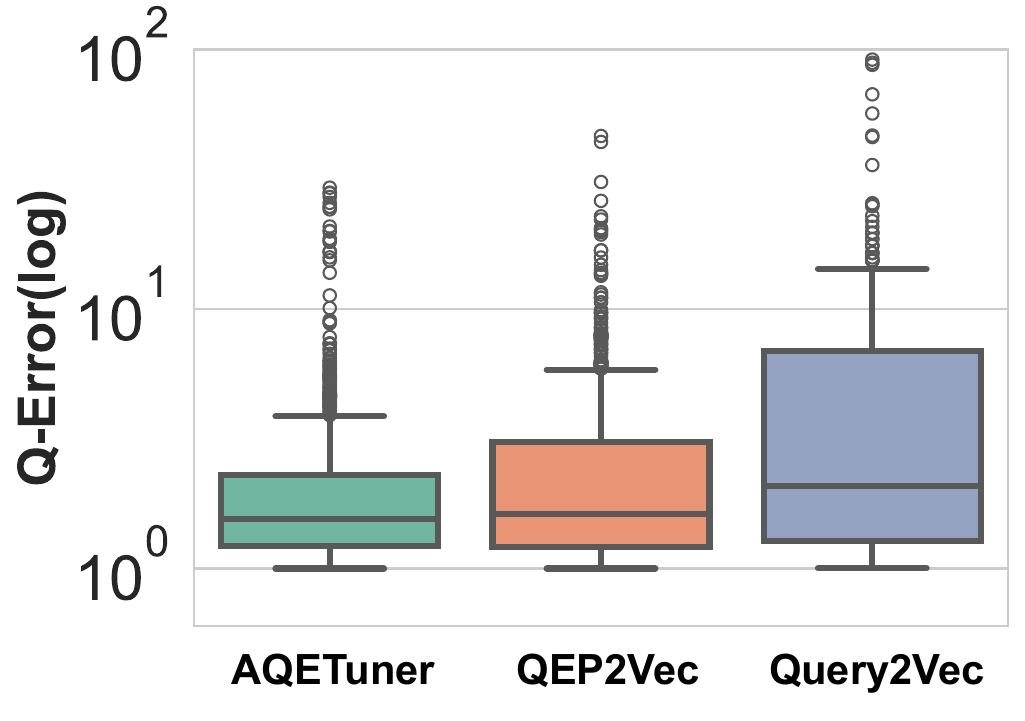}}}\hspace{1mm}
    \subfloat[{ROC Curve of Query Reliability}]{
    \label{fig:encoding_method_roc}
    {\includegraphics[width=0.47\linewidth]{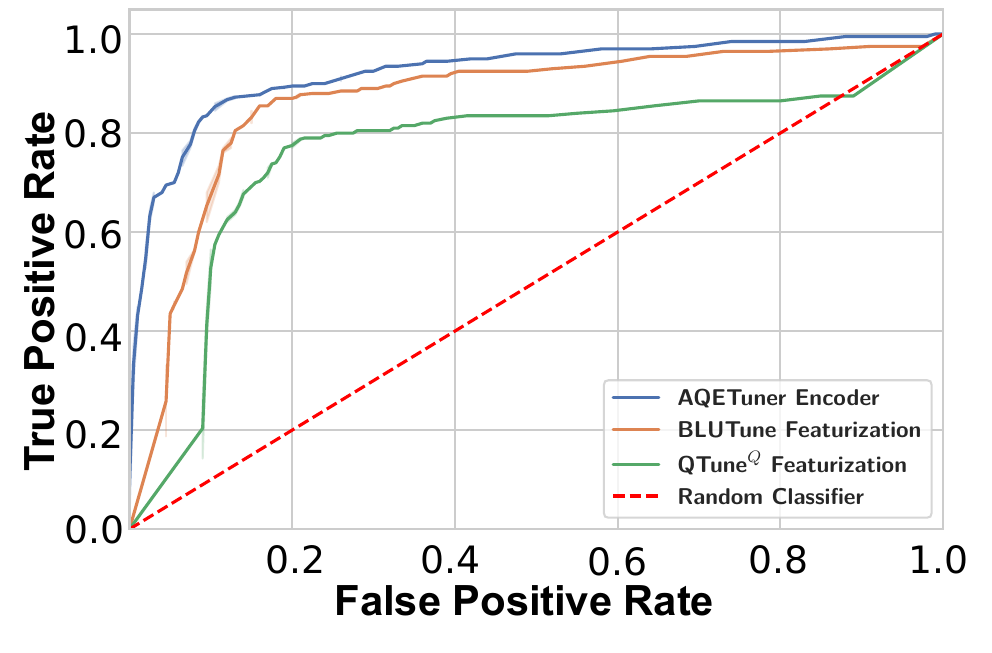}}}
  
  \label{fig: encoding_methods_effect}
  \vspace{-1em}
  \caption{Impact Evaluation of Knob-plan Encoders.}
  \vspace{-1em}
\end{figure}

To evaluate the effectiveness of different encoders, we apply them to two downstream prediction tasks: query performance and reliability.
Each encoder is followed by the same MLP layer to produce the final prediction.
To ensure a fair comparison, we concatenate the
query/plan
encodings from Query2Vec and QEP2Vec with \AQETune's self-attention-based knob encoding, forming the feature vectors for knob-plan pairs.
The evaluation datasets are obtained from the tuning observations of Section~7.1, which contained the true prediction targets.
For predict evaluation, we use the Q-Error metric~\cite{moerkotte2009preventing}, which measures the prediction error as: Q-Error(t) = $\max$($\frac{actual(t)}{predicted(t)}$, $\frac{predicted(t)}{actual(t)}$), where $t$ is the execution time.
For reliability evaluation, we use the ROC curve, a common tool for assessing classification models.

Figure~\ref{fig:encoding_method_box_data} shows the box plot of Q-Error distributions for performance predictions.
The \AQETune Encoder achieves the lowest median error and a more compact interquartile range (IQR), indicating more consistent and accurate performance predictions.
As shown in the ROC curve in Fig.~\ref{fig:encoding_method_roc}, the red dashed line indicates the baseline performance of a random classifier, with all three methods performing substantially better.
The curve of \AQETune Encoder is closest to the upper left corner,
which indicates that it has the highest prediction accuracy regarding query failures. 

In both prediction tasks, Query2Vec has the worst performance as it only extracts keywords and cost information from the query plan without considering its structure. QEP2Vec improves it by incorporating plan structures using fixed templates, such as predefined multi-way joins. 
However, it still struggles to capture the  complex patterns present in analytical queries.
In contrast, \AQETune's encoder utilizes fined-grained featurizations and attention mechanisms to capture the structural information of the query plans and knob-plan correlations, both of which are critical to the two downstream prediction tasks.
The experimental results show that \AQETune's encoder is more effective than other methods.


\begin{table}[t]
\caption{Evaluation of Surrogate Models.}
\vspace{-1.5em}
\scalebox{0.7}{
\begin{tabular}{c|c|c|c} 
\hline\hline
\rowcolor{mygrey}
\textbf{Surrogate Model} & \begin{tabular}[c]{@{}l@{}} \textbf{Avg. Infer Time (s)} \end{tabular}  & \textbf{\# of Observations}  & \textbf{RMSE} \\ \midrule
GP &  2.29 &  2827  & 3.14 \\ \midrule
CGP &  2.10  &  2842  & 3.49 \\ \midrule
DTP & \textbf{0.58}  & \textbf{3831}  & \textbf{2.96} \\ \bottomrule
\end{tabular}
}
\vspace{-1.5em}
\label{tab: surrogates}
\end{table}




\vspace{-2mm}
\subsection{Effect of Dual-Task Predictor}
To evaluate the effectiveness of the dual-task predictor (DTP) in \AQETune, we compare it with two alternative models: Gaussian Process (GP) and Contextual Gaussian Process (CGP)~\cite{krause2011contextual}.
GP uses an RBF kernel to determine the similarity between two points in the input space, while CGP incorporates contextual information (i.e., query plans) with a composite kernel.
Since GP and CGP can only handle a single prediction objective, we set their objective to query performance.
For a fair comparison, we use the cross-attention-based knob-plan encoding from \AQETune's encoder as input to GP. 
For CGP, which can capture the relationship between plans and knobs through its composite kernel, we use only the self-attention-based plan encoding as its input.



Over a fixed 6-hour tuning duration, we evaluate different surrogate models based on their average inference time, the number of collected observations, and the RMSE of prediction for execution time on the JOB-Extend workload.
As shown in Table~\ref{tab: surrogates},
we observe that DTP demonstrates much faster inference times than GP and CGP. 
This is because GP and CGP need to compute the covariance matrix with cubic complexity regarding the number of observations.
Besides, during the fixed duration, DTP could collect more observations than GP and CGP due to its lightweight computation and less knob recommendation time. 
Regarding the RMSE, DTP yields the lowest error compared to GP and CGP. It effectively leverages observation data on query reliability and exploits its correlation with query performance to enhance prediction accuracy.

\begin{figure}[t]
\centering
\subfloat[Diversity Analysis.]{
\label{fig: warm_start_diversity}
{\includegraphics[scale=0.25]{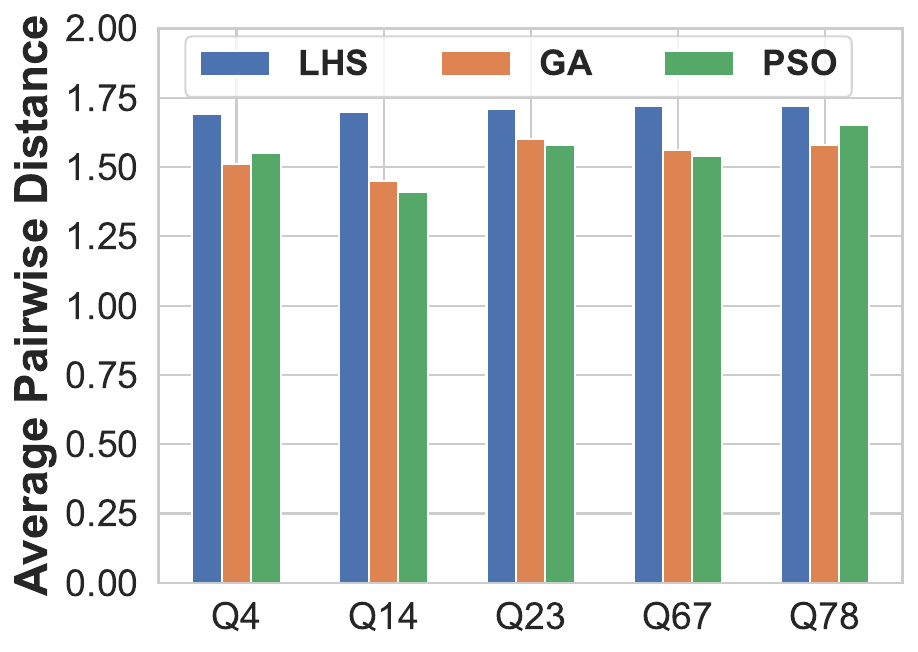}} 
}
\hspace{1mm}
\subfloat[Sample Efficiency Analysis.]{
    \label{fig: warm_start}
    {\includegraphics[scale=0.23]{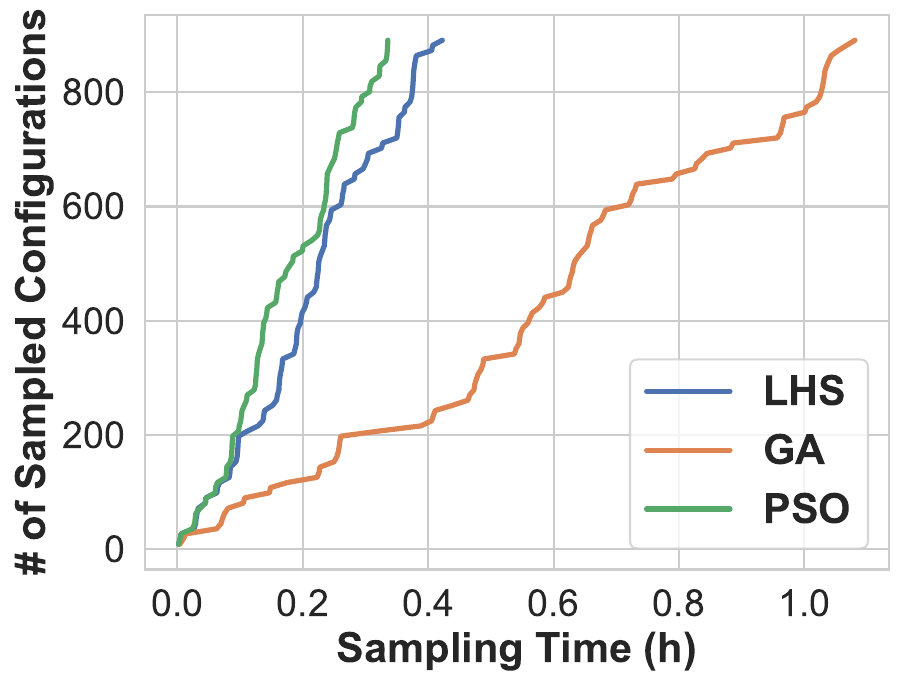}} }

  \vspace{-1em}
  
  \caption{Evaluation of Different Warm Starters.}
  \vspace{-1.5em}
\end{figure}

\vspace{-2mm}
\subsection{Assessment of Warm Starter}
To demonstrate the effectiveness of the PSO-based warm-starter, we compare it with two methods: Latin Hypercube Sampling (LHS) and Genetic Algorithm (GA).
LHS is commonly used in BO-based tuning methods to generate initial samples~\cite{zhang2021restune, van2017ottertune, zhang2022onlinetune}, 
while GA is employed by Hunter~\cite{cai2022hunter} to address the cold-start issue.
The comparison focuses on two criteria: \textit{diversity} and \textit{efficiency}.
For \textit{diversity}, we calculate the average pairwise Euclidean distance between sampling points in the knob space of each method.
We selected TPC-DS queries Q4, Q14, Q23, Q67, and Q78 for evaluation due to 
their complex query structures.
As shown in Fig.~\ref{fig: warm_start_diversity},
we observe that LHS achieves the largest distance, indicating the highest diversity. This is because LHS is a near-random sampling method that ensures the knob space is fully explored.
GA- and PSO-based methods show smaller average distances between their sample points as they tend to focus on potentially good regions.
However, the distance gap between these methods is not significant.

To measure the sample \textit{efficiency}, we compare how the number of samples increases over time for different methods, as shown in Fig.~\ref{fig: warm_start}.
All the methods terminate after finding 900 sample points.
We observe that PSO requires 20.6\% less time than LHS and 68.9\% less time than GA to gather sufficient samples.
While LHS performs well in maintaining sample diversity, it does not consider the tuning objective, leading to a longer execution time for its sample points than those from PSO, resulting in lower sampling efficiency. 
PSO is also much faster than GA because it allows parallel sampling, whereas GA requires the analytical engine to evaluate each point before the next one can be sampled. 

%% file: sections/related_work.tex
\vspace{-1mm}
\section{Related Work}

\noindent\textbf{DBMS Knob Tuning.}
To bypass the manual tuning process, 
various intelligent methods are proposed to automate knob tuning. 
They can be grouped into three categories:
(i) \textit{Heuristic methods} select configurations using predefined rules. BestConfig~\cite{zhu2017bestconfig} utilizes heuristics to tune within a resource constraint. Wei et al.~\cite{wei2014self} create fuzzy rules to capture relationships between configurations. However, these hard-coded rules often fail to leverage knowledge from past tuning.
(ii) \textit{RL-based methods} use reinforcement learning to recommend configurations by training a neural network model to capture the relationship between system metrics and knobs. CDBTune~\cite{zhang2019cdbtune} and Hunter~\cite{cai2022hunter} use DDPG to find optimal configurations in high-dimensional spaces. QTune~\cite{li2019qtune} uses a Double-State DDPG model for multi-level query-aware tuning. WATuning~\cite{ge2021watuning} uses attention-based deep RL to adapt to dynamic workloads.
(iii) \textit{BO-based methods} like iTuned~\cite{duan2009itune} and OtterTune~\cite{van2017ottertune} adopt Bayesian Optimization, building a surrogate model to balance exploration and exploitation.  ResTune~\cite{zhang2021restune} adds constraints to optimize resource usage without exceeding performance limits.  
Most of these methods focus on system-level tuning and do not fully address the specific needs of query-level tuning.

\noindent\textbf{Contextual Bayesian Optimization.}
Contextual Bayesian Optimization (CBO) extends the traditional BO framework to account for various objective functions influenced by contextual variables, making it particularly effective in dynamic environments. 
CBO typically uses Contextual Gaussian Processes (CGP)~\cite{krause2011contextual} as the surrogate model, combining kernels for both contexts and optimization actions.
This allows the optimization process to adapt to changing conditions, such as user preferences or system states, enhancing the robustness of BO in real-world applications. 
A neural network-assisted Gaussian Process model has also been developed~\cite{zhang2024contextual} to approximate the target function based on context variables.
However, CGP still faces high computational costs. In our work, we use a Neural Process-based dual-task predictor to lower the computation overhead while maintaining a comparable prediction accuracy.

%% file: sections/conclusion.tex
\vspace{-1mm}
\section{Conclusion}

This paper presents \AQETune, a BO-based tuning approach for reliable query-level tuning in analytical query engines. 
It utilizes a knob-plan encoder to capture the correlation between tunable knobs and query plans and a dual-task model to predict query performance and reliability.
Besides, it employs a PSO-based warm-starter to address the cold-start issue. 
\AQETune achieves the best results for various benchmarks compared to the SOTA methods.